\documentclass[aps,showpacs,preprint,epsf,epsfig, showkeys, nofootinbib]{revtex4-2}
\usepackage{mathrsfs}
\usepackage{dcolumn}
\usepackage{bm}
\usepackage{hyperref}
\usepackage{epsfig,graphicx,color,appendix}
\usepackage{multirow}
\usepackage{tabularx}
\usepackage{capt-of}
\usepackage{caption}
\usepackage[countmax]{subfloat}

\usepackage{threeparttable}
\usepackage{amssymb}
\usepackage{array}
\usepackage{amsmath}
\usepackage{rotating,tabularx}
\usepackage{float}
\usepackage[english]{babel}

\begin{document}

\title{Systematic study of light and charm meson M1 radiative transitions}
\author{\textbf{Binesh Mohan}}
\email{bineshmohan96@gmail.com}
\author{\textbf{Christas \surname{Mony A.}}}
\email{christasmony@gmail.com}
\author{\textbf{Rohit Dhir}}
\email[Corresponding author: ]{dhir.rohit@gmail.com}
\affiliation{Department of Physics and Nanotechnology,\\SRM Institute of Science and Technology, Kattankulathur 603203, India.}
	
\long\def\symbolfootnote[#1]#2{\begingroup%
\def\thefootnote{\fnsymbol{footnote}}\footnote[#1]{#2}\endgroup}
\def\lsim{ {\ \lower-1.2pt\vbox{\hbox{\rlap{$<$}\lower5pt\vbox{\hbox{$\sim$}
        }}}\ } }
\def\gsim{ {\ \lower-1.2pt\vbox{\hbox{\rlap{$>$}\lower5pt\vbox{\hbox{$\sim$}
        }}}\ } }

\vskip 1.5 cm
\small
\vskip 1.0 cm

\date{\today}
\begin{abstract}
Motivated by recent experimental advancements in the study of radiative decays of charmed mesons, we investigate the magnetic (transition) moments of vector mesons by applying the effective mass scheme, obtained from the one-gluon exchange interaction between quark-antiquark pairs. By incorporating high-precision experimental data from both heavy and light flavor sectors, we accurately account for the strong hyperfine interaction contributions to the quark and antiquark masses within mesons. We calculate the $V \to P$ effective transition magnetic moments to reliably predict M1 decay widths. Furthermore, to enhance the completeness of our analysis, we employ the non-relativistic potential model to calculate bound state isomultiplet masses and to predict the M1 decay widths of charmed mesons. Additionally, scale-dependent effects in both the effective mass scheme and the potential model are systematically analyzed and interpreted, emphasizing the decisive role of higher-order QCD corrections in determining M1 decay widths.
\end{abstract}

\maketitle
\newpage
	
\section{Introduction}	
\label{Intro}
The past decade has seen a surge in experimental activity focused on the masses, properties, and decays of heavy-flavor hadrons. However, most hadronic states are unstable against strong interactions, resulting in extremely short lifetimes (on the order of $10^{-23}$ s). Consequently, their magnetic moments are exceptionally challenging to measure directly using conventional methods due to their fleeting existence. Nevertheless, indirect estimates of the $\rho^+$ meson's magnetic moment, $\mu_{\rho^{+}} = 2.6(6)~\mu_N$\footnote{Note that values in the parentheses represent uncertainties.}, have been derived from cross-section measurements of the $e^{+}e^{-}\to\pi^{+}\pi^{-}2\pi^{0}$ process by the BABAR Collaboration \cite{Druzhinin:2007cs, GarciaGudino:2015ocw, BaBar:2017zmc}. This result was consistent with various theoretical models but suffered from significant uncertainty. More recently, an improved precision estimate yielded a larger value of $\mu_{\rho^{+}} = 3.3(4)~\mu_N$ \cite{Rojas:2024tmn}, falling outside the range predicted by most theoretical approaches. Beyond magnetic moments, analyzing M1 radiative decay widths (transition magnetic moments and form factors) offers a complementary approach to investigate hadron structure. These observables serve as sensitive probes of quark-gluon dynamics within bound states. In particular, M1 transitions provide a powerful testing ground for QCD-based theoretical models, as their electromagnetic decay widths directly reflect the wave function overlap and spin-flip dynamics of constituent quarks. Presently, experimental data for vector ($V$) to pseudoscalar ($P$) M1 radiative decay widths are available for light mesons, as well as for the charm mesons $D^{*+} \to D^+\gamma$ and $J/\psi \to \eta_c\gamma$ \cite{ParticleDataGroup:2024cfk}. However, only upper limits exist for the M1 decay widths of $D^{*0} \to D^0\gamma$ and $D_s^{*+} \to D_s^+\gamma$ due to insufficient precision in their total width measurements \cite{CLEO:1995ewe, Abachi:1988fw}. Recent advancements include the BESIII Collaboration's measurement of $D^{*0}$ branching fractions: $\mathcal{B}(D^{*0} \to D^0\pi^0) = (65.5\pm 0.8\pm 0.5)\%$ and $\mathcal{B}(D^{*0} \to D^0\gamma) = (34.5\pm 0.8\pm 0.5)\%$, assuming only these two decay channels \cite{BESIII:2014rqs}. More recently, BESIII reported branching fractions for $D_s^{*+} \to D_s^+\gamma$ and $D_s^{*+} \to D_s^+\pi^0$ as $(93.57\pm 0.38\pm 0.22)\%$ and $(5.76\pm 0.38\pm 0.16)\%$, respectively, utilizing the world average for $D_s^{*+} \to D_s^+e^+e^-$ \cite{BESIII:2022kbd}. While $B^{*} \to B\gamma$ and $B_s^{*} \to B_s\gamma$ radiative M1 decay processes have been observed in the bottom meson sector, concrete experimental data for their decay widths remain elusive to date \cite{ParticleDataGroup:2024cfk}. Further experimental investigations in this direction are anticipated from collaborations such as LHCb and BESIII in the near future.

Theoretical investigations into meson magnetic moments and radiative M1 decays have been extensively carried out using various models, including extended bag model (EBM) \cite{Simonis:2018rld, Simonis:2016pnh}, light cone QCD sum rules (LCQSR) \cite{Pullin:2021ebn, Aliev:2019lsd, Aliev:2009gj}, MIT bag model (MIT BM) \cite{Zhang:2021yul}, Nambu-Jona-Lasinio (NJL) model \cite{Luan:2015goa, Deng:2013uca}, model based on Dyson-Schwinger/Bethe-Salpeter equations (DSE/BSE) \cite{Xu:2024fun, Bhagwat:2006pu, Maris:2002mz}, relativistic Hamiltonian (RH) \cite{Badalian:2012ft}, lattice QCD (LQCD) \cite{Meng:2024gpd, Colquhoun:2023zbc, Owen:2015gva, Becirevic:2009xp, Lee:2008qf}, chiral perturbation theory ($\chi$PT) \cite{Meng:2022ozq, Wang:2019mhm}, relativistic potential model (RPM) \cite{Jena:2010zza, Jena:2002is}, constituent quark model (CQM) \cite{Wang:2023bek}, light-front holographic model (LFHM) \cite{Ahmady:2020mht}, light-front quark model (LFQM) \cite{Choi:2007se, Choi:1997iq}, QCD sum rules (QCDSR) \cite{Lu:2024tgy}, potential model (PM) \cite{Bonnaz:2001aj}, relativistic independent quark (RIQ) model \cite{Priyadarsini:2016tiu}, relativistic quark model (RQM) \cite{Ebert:2002xz, Ebert:2002pp}, covariant model (CM) \cite{Cheung:2015rya, Cheung:2014cka}, Godfrey-Isgur (GI) model \cite{Godfrey:2015dva, Godfrey:1985xj}, and covariant confined quark model (CCQM) \cite{Tran:2023hrn}. Despite these efforts, a unified and consistent theoretical framework that accurately describes radiative M1 decays across both light- and heavy-flavor mesons remains elusive.

The effective mass scheme (EMS) has previously demonstrated success in studying the magnetic properties of baryons \cite{Mohan:2022sxm, Hazra:2021lpa, Dhir:2013nka}. In a fresh attempt, we systematically extend the EMS framework to investigate meson bound states, anticipating comparable effectiveness. In this work, we employ the EMS to calculate the magnetic moments of vector mesons, utilizing effective quark masses derived from experimental meson masses reported by the Particle Data Group (PDG) \cite{ParticleDataGroup:2024cfk}. We then evaluate the corresponding $V \to P$ transition moments to make robust predictions for radiative M1 decays. While our primary focus is on charm mesons, we also include results for light mesons to fine-tune our approach against existing experimental data. Additionally, for charm mesons, we utilize a non-relativistic potential model (PM) to compute bound state masses and spatial wave functions, subsequently predicting their M1 decay widths. Crucially, unlike the spin-flavor wave functions of EMS, the PM provides numerical estimates of spatial wave functions, offering a unique comparative scope. The PM's wave functions, influenced by phenomenological parameters, exhibit flavor and scale-dependence. Given that M1 decay matrix elements are sensitive to these wave functions, their detailed study can yield vital insights into these processes and significantly improve predictions. We strengthen the theoretical foundation of our study by first employing the PM, where anomalous magnetic moment contributions, relativistic corrections, and scale-dependent effects are systematically incorporated. By explicitly accounting for the sensitivity of the strong coupling constant and the wave function to the renormalization scale, we achieve a substantial improvement in the description of M1 transitions within this framework. Extending the analysis to the EMS, we further interpret the magnetic structure of mesons through a scale-dependent treatment, thereby resolving ambiguities in M1 transition widths. Crucially, we argue that the observed scale-dependence reflects genuine QCD corrections beyond the static quark model, leading to a unified and more accurate description of M1 transitions across the charm sector. This dual approach, employing EMS and PM, thus provides a comprehensive perspective on the radiative M1 transitions of heavy meson bound states.

The paper is organized as follows: Secs. \ref{EMS} and \ref{Properties} present the EMS methodology, while Sec. \ref{Potential} describes the PM framework. Sec. \ref{Discussions} contains our numerical results and compares them with existing theoretical models. Finally, Sec. \ref{Conclusions} summarizes our key findings and conclusions.
\section{Effective Mass Scheme for Mesons}
\label{EMS}
The mass of a quark inside a hadron can be modified due to the one-gluon exchange interaction with the spectator quarks and is referred to as the effective quark mass in the concept of EMS \cite{DeRujula:1975qlm}. According to EMS, the meson mass, $M_\mathcal{M}$, is calculated as the sum of effective masses of its constituents (quark and antiquark), and it can also be written as the sum of constituent quark masses and the spin-dependent strong hyperfine interaction term between them \cite{Mohan:2022sxm, Hazra:2021lpa, Dhir:2013nka}. Thus,
\begin{equation}
\label{eq1} 
    {M}_\mathcal{M} = m_{q}^{\mathscr{E}} + m_{\overline{q}^\prime}^{\mathscr{E}} = {m_q} + {m_{\overline{q}^\prime}} + b_{q\overline{q}^\prime}~\bf{s_q.s_{\overline{q}^\prime}},
\end{equation}
where $m_{q}^{\mathscr{E}} (m_{\overline{q}^\prime}^{\mathscr{E}})$ and $m_{q} (m_{\overline{q}^\prime})$ represent the effective and constituent masses of the quark (antiquark) within the meson, respectively; and the corresponding spin operators of the quark and antiquark are denoted by $\bf{s_q}$ and $\bf{s_{\overline{q}^\prime}}$. The strong hyperfine interaction term\footnote{Note that, $b_{q\overline{q}^\prime}=b_{\overline{q}^\prime q}$.}, $b_{q\overline{q}^\prime}$, for meson $\mathcal{M}(q\overline{q}^\prime)$ is given by \cite{DeRujula:1975qlm}
\begin{equation}
\label{eq2}
b_{q\overline{q}^\prime} =\frac{32\pi\alpha_s}{9m_qm_{\overline{q}^\prime}}\langle\psi|\delta^3(\vec{r})|\psi\rangle,
\end{equation}
where $\psi$ is the meson wave function at the origin and $\alpha_s$ is the strong coupling constant. The mass of a quark inside a meson, $\mathcal{M}(q\overline{q}^\prime)$, changes due to its interaction with an antiquark. Therefore, general expressions for meson masses can be written in terms of effective quark and antiquark masses as:
\begin{enumerate}	
	\item For pseudoscalar ($J^P=0^-$) mesons,
		\begin{equation}
			\label{eq3}
			M_{P}  = m_{q} + m_{\overline{q}^\prime} - \frac{3b_{q\overline{q}^\prime}}{4}. 
		\end{equation}
	\item For vector ($J^P=1^-$) mesons,
	\begin{equation}
		\label{eq4}
		M_{V}  = m_{q} + m_{\overline{q}^\prime} + \frac{b_{q\overline{q}^\prime}}{4},
	\end{equation}
\end{enumerate}
where $q$ and $\overline{q}^\prime$ represent the quark flavors \textit{u, d, s}, and \textit{c}. For a pseudoscalar meson, the effective masses for the quark and antiquark can be defined as,
\begin{equation}
\begin{split}
\label{eq5}
  m_{q}^{\mathscr{E}} = m_{q} + \alpha b_{q\overline{q}^\prime}, \\
  m_{\bar{q \prime}}^{\mathscr{E}} = m_{\overline{q}^\prime} + \alpha b_{q\overline{q}^\prime}.
  \end{split}
\end{equation}
Following Eq. \eqref{eq1}, the resultant mass expression for the pseudoscalar meson can be written as, 
\begin{equation}
\label{eq6}
   M_{P} = m_{q} + m_{\overline{q}^\prime} + 2\alpha b_{q\overline{q}^\prime}.
\end{equation}
The $\alpha$ parameter is determined by comparing Eq. \eqref{eq6} with the mass expression of the pseudoscalar meson given in Eq. \eqref{eq3}. Thus, we obtain,
\[ \alpha = -\frac{3}{8}. \]
Therefore, the general expressions for effective quark and antiquark masses in pseudoscalar meson is given by
\begin{equation}
\begin{split}
\label{eq7}
  m_{q}^{\mathscr{E}} = m_{q} - \frac{3b_{q\overline{q}^\prime}}{8}, \\
  m_{\bar{q \prime}}^{\mathscr{E}} = m_{\overline{q}^\prime} - \frac{3b_{q\overline{q}^\prime}}{8}.
  \end{split}
\end{equation}
In a similar way, the effective masses for quark and antiquark in a vector meson is obtained as,
\begin{equation}
\begin{split}
\label{eq8}
  m_{q}^{\mathscr{E}} = m_{q} + \frac{b_{q\overline{q}^\prime}}{8}, \\
  m_{\bar{q \prime}}^{\mathscr{E}} = m_{\overline{q}^\prime} + \frac{b_{q\overline{q}^\prime}}{8}.
  \end{split}
\end{equation}

The effective quark mass $m_q^{\mathscr{E}}$ in EMS is closely analogous to the leading-order parametrization of meson masses in $\chi$PT \cite{Morpurgo:1989ti, Dillon:1995qw}, but extends beyond leading-order through the explicit $1/(m_q m_{\overline{q}^\prime})$ dependence of the hyperfine term. This structure is consistent with the broader QCD picture, where flavor breaking in the Lagrangian originates from quark mass differences and electromagnetic charges are carried solely by quarks \cite{Dillon:2002ks}. Higher-order nonlinear corrections are suppressed, making the nonrelativistic quark model (NRQM) parametrization of masses and magnetic moments effectively equivalent to leading-order relativistic field-theoretical treatments \cite{Durand:2001zz, Durand:2001sz, Dillon:2002ks}. Within EMS, constituent quark masses act as renormalized parameters that already incorporate confinement, chiral symmetry breaking, and non-hyperfine gluonic effects, leaving the hyperfine interaction as the explicit residual force. As emphasized by Morpurgo \textit{et al.} \cite{Morpurgo:1999mr}, the general formulas for meson masses (Eqs. \eqref{eq3} and \eqref{eq4}) thus reproduce spectra through first-order flavor breaking in $m_q$ and hyperfine terms $b_{q\overline{q}^\prime}$. Using PDG data \cite{ParticleDataGroup:2024cfk}, we determine $m_q$, $m_{\overline{q}^\prime}$, and $b_{q\overline{q}^\prime}$ (Table \ref{t1}), ensuring parameter-independent predictions of meson masses and magnetic properties. In this framework, $b_{q\overline{q}^\prime}$ absorbs the combined scale-dependence of $\alpha_s$ and $|\psi(0)|^2$, while subdominant electromagnetic and isospin effects are effectively included in the empirical effective masses.  

A key challenge in constituent models is that $\alpha_s$ and $|\psi(0)|^2$ are both scale-dependent and strongly correlated, obscuring their separate roles in magnetic transitions. Conventional approaches often retune $\alpha_s$ or rescale wave functions across meson families, reducing predictivity. Here we refine EMS by treating $\alpha_s$ and $|\psi(0)|^2$ jointly: fixing $\alpha_s$ in the well-measured charmonium sector and extracting effective $|\psi(0)|^2$ values from hyperfine splittings (Eq. \eqref{eq2}) across flavors. The variation of wave function overlap with meson size then emerges self-consistently through $b_{q\overline{q}^\prime}$, embedding QCD scale-dependence directly into EMS. This elevates EMS into a self-consistent and parameter-lean framework, where the interplay between $\alpha_s$ and $|\psi(0)|^2$ is absorbed into effective observables. The scheme thus provides fresh insight into the magnetic structure of mesons and resolves the ambiguity in radiative transitions. In later sections, we further analyze the role of $\alpha_s$ through the anomalous magnetic moment factor $(1+\kappa_Q)$ in M1 decays of heavy flavors.
\begin{table}[ht]
\centering
\captionof{table}{Constituent quark masses and hyperfine interaction terms (in GeV).}
\label{t1}
\setlength{\tabcolsep}{2pt}
\begin{tabular}{|c|c||c|c|} 	\hline 
	\textbf{Experimental}			&\textbf{Constituent Quark} & \textbf{Experimental} &\textbf{Hyperfine Interaction} \\
	\textbf{Inputs \cite{ParticleDataGroup:2024cfk}} &\textbf{Masses ($m_{q} (m_{\overline{q}^\prime})$)} & \textbf{Inputs \cite{ParticleDataGroup:2024cfk} } &\textbf{Terms ($b_{q\overline{q}^\prime}$)}\\	\hline \hline
    $\rho^{+}, \pi^{+}$ & $m_{u} = m_{d} =0.306$ & $\rho^{+}, \pi^{+}$     & $b_{u\bar{u}} = b_{u\bar{d}} (= b_{d\bar{u}}) = b_{d\bar{d}} = 0.629$ \\ \hline
    $K^{*+}, K^{+}$     & $m_{s} = 0.489$ & $K^{*+}, K^{+}$         & $b_{u\bar{s}} (= b_{s\bar{u}}) = b_{d\bar{s}} (= b_{s\bar{d}}) = 0.402$ \\
	&                                       & $\phi$                  & $b_{s\bar{s}} = 0.164$  \\	\hline	
	$D_{s}^{*+}$        & $m_{c} = 1.587$ & $D^{*0}$, $D^{0}$       & $b_{u\bar{c}} (= b_{c\bar{u}}) = b_{d\bar{c}} (= b_{c\bar{d}}) = 0.142$  \\
	&                                       & $D_{s}^{*+}, D_{s}^{+}$ & $b_{s\bar{c}} (= b_{c\bar{s}}) = 0.144$ \\
	&                                       & $J/\psi$, $\eta_c$      & $b_{c\bar{c}} = 0.113$ \\	\hline 
\end{tabular} 
\end{table}   

We wish to remark that $\eta-\eta^{\prime}$ and $\omega-\phi$ states are also studied in our current work. The mixing scheme used for $\eta$ and $\eta^{\prime}$ states is given by
\begin{equation} \label{eq9}
\begin{split}
    \eta^{\prime} &= \frac{1}{\sqrt{2}}(u\bar u + d\bar d)\cos{\phi_P} + (s\bar s)\sin{\phi_P},\\
    \eta &= \frac{1}{\sqrt{2}}(u\bar u + d\bar d)\sin{\phi_P} - (s\bar s)\cos{\phi_P},
\end{split} 
\end{equation}
where $\phi_P = \theta_{ideal} - \theta_P^{phys}$ and $\theta_P^{phys} = -14.1^{\circ}$ \cite{ParticleDataGroup:2024cfk}. For $\omega$ and $\phi$ vector states, ideal mixing is assumed, \textit{i.e.}, $\omega = \frac{1}{\sqrt{2}}(u\bar u + d\bar d)$ and $\phi = s\bar s$. 

\section{Magnetic Properties of Mesons}
\label{Properties}
The magnetic moments of the ground state vector mesons, $\mu_{V}$, can be obtained by sandwiching the magnetic moment operator, $\pmb{\mu}$, between the corresponding meson wave functions as given below.
\begin{equation} 
\label{eq10}  
    \mu_{V} = \langle\Psi_{sf}|\pmb{\mu}|\Psi_{sf}\rangle,
\end{equation}
where $|\Psi_{sf}\rangle$ is the spin-flavor wave function of the corresponding meson state. The theoretical magnetic moment operator, $\pmb{\mu}$, is given by
\begin{equation} 
\label{eq11}
    \pmb{\mu} =\sum_{i}\mu_i^{\mathscr{E}} \pmb{\sigma_{i}},  
\end{equation}
where $\pmb{\sigma_{i}}$ denote the Pauli spin matrices and $\mu_i^{\mathscr{E}}$ represent the effective quark magnetic moment given as
\begin{equation} \label{eq12}
    \mu_i^{\mathscr{E}} =\sum_{i}\frac{e_{i}}{2m_{i}^{\mathscr{E}}}, 
\end{equation}
where \textit{i = u, d, s}, and \textit{c}; $m_{i}^{\mathscr{E}}$ represent the effective mass of quark $i$ with a bare charge $e_{i}$. The low-energy photon used for probing magnetic moments can observe only the overall internal structure of the meson. Thus, the magnetic moment of a meson state can be expressed in terms of the effective quark (antiquark) magnetic moments, which contain a mass term that is interpreted as the effective mass of the quark. Therefore, the general expression for the magnetic moment of vector mesons is given as follows:
\begin{equation} 
\label{eq13}
    \mu_{V} = \mu_{q}^{\mathscr{E}} + \mu_{\overline{q}^\prime}^{\mathscr{E}}.
\end{equation}
Further, we calculate the transition magnetic moments for $V\to P$ transition by sandwiching Eq. \eqref{eq11} between the initial and final state wave functions of vector and pseudoscalar meson states. We obtain the following general expression for the transition magnetic moment corresponding to the $V\to P$ transition.
 \begin{equation} \label{eq14}
 	\mu_{V \to P} = \mu_{q}^{\mathscr{E}} - \mu_{\overline{q}^\prime}^{\mathscr{E}}.
 \end{equation}
In the evaluation of transition magnetic moments, we utilize the quark transition mass defined as the arithmetic mean of effective masses of constituent quarks (antiquarks) of initial and final state mesons. The transition magnetic moments thus obtained are utilized to calculate the radiative M1 decay widths corresponding to $V\to P\gamma$ decays using the given relation \cite{Verma:1980fd}.
\begin{equation} \label{eq15}
	\Gamma_{V\to P\gamma} = {\frac{1}{12\pi}}{\omega^3}|{\mu_{V\to P}}|^{2},
\end{equation}
where
\begin{equation} \label{eq16}
	\omega = \frac{M^2_{V} - M^2_{P}} {2M_{V}}
\end{equation}
is the photon momentum in the rest frame of the decaying particle, and $\mu_{V\to P}$ is the transition magnetic moment expressed in nuclear magneton, $\mu_N$. Here, $M_V$ and $M_P$ are the masses of vector and pseudoscalar mesons, respectively, which determine the momentum carried by the photon. As pointed out in our recent works, reliable prediction of radiative M1 decay widths requires the precise evaluation of photon momenta. To achieve this, we utilized the experimental masses of meson states available in PDG \cite{ParticleDataGroup:2024cfk}. We have discussed and listed the photon momenta involved in $V \to P$ transitions in Sec. \ref{Discussions}. 

Theoretically, our framework provides a consistent formalism for evaluating both the static magnetic moments of vector mesons and the transition magnetic moments for $V \to P\gamma$ decays. The introduction of the quark transition mass, defined as the arithmetic mean of the effective quark masses, captures the dynamical behavior associated with the off-shell nature of quarks during photon emission, thereby moving beyond naive treatments. The transition magnetic moment $\mu_{V\to P}$ emerges as the key quantity linking spin–magnetic dynamics to observable radiative processes. Experimentally, M1 transitions such as $J/\psi \to \eta_c \gamma$ and $D^{*}\to D\gamma$ are well measured and provide stringent tests of quark model predictions. Hence, a reliable determination of $\mu_{V\to P}$ and its scale-dependence is essential for reconciling theoretical descriptions with data and for quantifying the impact of relativistic and QCD corrections. 
\section{Potential Model Approach}
\label{Potential}
The charmed, charm-strange, and charmonium bound states can also be studied in a non-relativistic potential model approach; we provide a brief description of PM in this section. The interquark potential is governed by a Lorentz vector color Coulombic one-gluon exchange interaction at short-range and a Lorentz scalar linear confinement interaction at long-range. Accordingly, we employ the Cornell potential given as \cite{Eichten:1974af, Eichten:1979ms}:
\begin{equation}
\label{eq17}
    V(r) = -\frac{4\alpha_s}{3r} + \sigma r + V_c,
\end{equation}
where the phenomenological parameters $\alpha_s$ and $\sigma$ are the strong coupling constant and the string tension, respectively. The constant $V_c$ is a dimensional quantity corresponding to the normalization of the energy levels of the meson system.

The energy eigenvalue of the bound state is obtained by solving the Schrodinger wave equation, generically given as $[T + V] \Psi = E \Psi$, where the kinetic energy of the constituents is given by $T$ and employing the potential $V$ as given in Eq. \eqref{eq17}. Further, the total wave function $\Psi (\vec{r})$ can be factorized into radial and angular components, expressed in terms of the radial wave function $R_{nl}(r)$ and the spherical harmonics $Y_{lm}(\theta,\phi)$, \textit{i.e.}, $\Psi (\vec{r}) = R_{nl}(r) Y_{lm}(\theta,\phi)$. Thereby, we obtain the Schrodinger radial wave equation for a bound system given as
\begin{equation}
\label{eq18}
    u''_{nl}(r) + 2 \mu \bigg[ E_{nl} - V(r) - \frac{l(l+1)}{2\mu r^2} \bigg] u_{nl}(r) = 0,
\end{equation}
where the reduced radial wave function $u_{nl}(r) = r R_{nl}(r)$, with the normalization $\int_0^{\infty} |R_{nl}(r)|^2 r^2 dr = \int_0^{\infty} |u_{nl}(r)|^2 dr = 1$. $E_{nl}$ is the energy eigenvalue of the bound state and $\mu$ is the reduced mass of the quarks. We solve Eq. \eqref{eq18} numerically using the Runge-Kutta method to obtain $E_{nl}$, from which the degenerate mass of the bound state is calculated as $M = m_q + m_{\overline{q}^\prime} + E_{nl}$, where $m_q(m_{\overline{q}^\prime})$ is the mass of the quark (antiquark). Along with the energy eigenvalue, we also estimate the corresponding wave function of the bound state, which is crucial for the computation of radiative M1 transitions. Using the estimated radial wave function, its value at the origin can be determined through extrapolation. For $l=0$ states, the total wave function at origin is related to the radial wave function at origin by $\Psi (0) = \frac{R_{nS}(0)}{\sqrt{4\pi}}$.

The degeneracy in the mass of the bound state is broken through the perturbative inclusion of the spin-spin interaction, which contributes to hyperfine splitting \cite{Eichten:1979pu, Eichten:1980mw, Gromes:1984ma, Lucha:1995zv} and is given by 
\begin{equation}
\label{eq19}
    V_{SS}(r) = \frac{32\pi\alpha_s}{9m_q m_{\overline{q}^\prime}}\textbf{S}_{q}.\textbf{S}_{\overline{q}^\prime} \delta(r),
\end{equation}
with the analytical expression for the spin-spin operator as, $\langle \textbf{S}_{q}.\textbf{S}_{\overline{q}^\prime} \rangle = \frac{1}{2} s(s+1) - \frac{3}{4}$.
We choose the Dirac delta function, $\delta(r)$, to be given as the smeared Gaussian 
\begin{equation}
\label{eq20}
    \delta(r) = \Big(\frac{\rho}{\sqrt{\pi}}\Big)^{3} e^{-\rho^2r^2},
\end{equation}
which incorporates the relativistic correction of $\mathcal{O}(v^2/c^2)$ as stated in Ref. \cite{Barnes:2005pb}. The non-degenerate mass of the low-lying bound state is written as  
\begin{equation}
M_{P/V} = m_q + m_{\overline{q}^\prime} + E_{nl} + \langle u_{nl} |V_{SS}| u_{nl} \rangle,
\end{equation}
where the overlap is controlled by the smearing parameter $\rho$. The optimized potential model parameters employed to compute the pseudoscalar and vector meson masses and wave functions are summarized in Table~\ref{tab:2_Potentialparameters}. We fix $\alpha_s = 0.56$ uniformly for the charm sector, consistent with charmonium expectations~\cite{A:2023bxv}, and adopt $m_c = 1.5$~GeV along with $\sigma^{c\overline{c}}$, $V_c^{c\overline{c}}$, and $\rho^{c\overline{c}}$ from our earlier study~\cite{A:2023bxv}. The light-quark masses are chosen as $m_{u,d} = 0.325$~GeV and $m_s = 0.5$~GeV, within established ranges~\cite{Ebert:2002xz, Godfrey:2015dva, Jena:2010zza, Li:2010vx, Ni:2021pce, Hao:2022vwt, Hao:2024ptu}. The remaining parameters ($\sigma^{c\overline{u}}, \sigma^{c\overline{s}}, V_c^{c\overline{u}}, V_c^{c\overline{s}}, \rho^{c\overline{u}}, \rho^{c\overline{s}}$) are calibrated to reproduce the masses and hyperfine splittings of charm and charm-strange mesons, and found to be consistent with those for charmonium.

\begin{table}[h]
    \centering
\caption{Potential parameters for open and hidden charm systems.}
\label{tab:2_Potentialparameters}
    \begin{tabular}{|c|c|c|c|} \hline 
         \textbf{Parameters}&  $\mathbf{c\overline{u}}, \mathbf{c\overline{d}}$&  $\mathbf{c\overline{s}}$& $\mathbf{c\overline{c}}$ \cite{A:2023bxv}\\ \hline \hline 
         $\sigma$ in GeV${}^2$&  $0.114$&  $0.145$& $0.14$\\  
         $V_c$ in GeV&  $-0.25$&  $ -0.291$& $-0.014$\\  
         $\rho$ in GeV&  $1.776$&  $1.245$& $1.19$\\ \hline
    \end{tabular}
\end{table}

Finally, the decay width of the low-lying $V\to P\gamma$ decays of heavy flavored mesons in the non-relativistic approximation is given by \cite{Jackson:1976mt, Novikov:1977dq, Godfrey:2015dva}
\begin{equation} 
\label{eq21}
    \Gamma_{V\to P\gamma} = \frac{\alpha}{3} k_{\gamma}^3 \Big| \Big\langle f_P \bigg| \frac{e_q}{m_q} j_0 \bigg(k_{\gamma} r \frac{m_{\overline{q}^\prime}}{m_q + m_{\overline{q}^\prime}}\bigg) - \frac{e_{\overline{q}^\prime}}{m_{\overline{q}^\prime}} j_0 \bigg(k_{\gamma} r \frac{m_{q}}{m_q + m_{\overline{q}^\prime}}\bigg) \bigg| i_V \Big\rangle \Big|^2,
\end{equation}
where $i_V$ and $f_P$ are the initial and final state spatial wave functions, respectively. $j_0(x)$ is the spherical Bessel function. The photon momentum $k_{\gamma}$ is evaluated using the numerical values of the mass of states obtained from the potential model. $\alpha$ is the fine structure constant and $e_q(e_{\overline{q}^\prime})$ is the charge of the quark (antiquark). Subsequently, employing Eq. \eqref{eq21}, we give our predictions for the radiative M1 transitions in the next section.

It is important to note that, although Eq.~\eqref{eq21} is derived within a non-relativistic approximation, the decay width is intrinsically sensitive to the scale-dependence of QCD. Aforementioned, both the strong coupling constant $\alpha_s(\mu)$ and the wave function at the origin $|\psi(0)|^2_\mu$ (where $\mu$ denotes the renormalization scale) enter into the evaluation of hyperfine interactions and M1 matrix elements. Within the potential model, $|\Psi(0)|^2$ is obtained directly from the Schrödinger solution, while in the effective mass scheme the hyperfine term $b_{q\overline{q}^\prime}$ effectively absorbs the combined scale-dependence of $\alpha_s$ and $|\psi(0)|^2$. Consequently, different choices of $\mu$ or parametrizations can lead to non-negligible QCD corrections in the numerical predictions. We therefore emphasize that the numerical results in the Sec.~\ref{PM_m1} should be interpreted in light of this scale sensitivity, which we address systematically in Sec.~\ref{scale} by incorporating QCD corrections beyond the static quark model. 
\section{Numerical Results and Discussions}
\label{Discussions}
\begin{figure}[]
    \centering
    \includegraphics[width=0.95\linewidth]{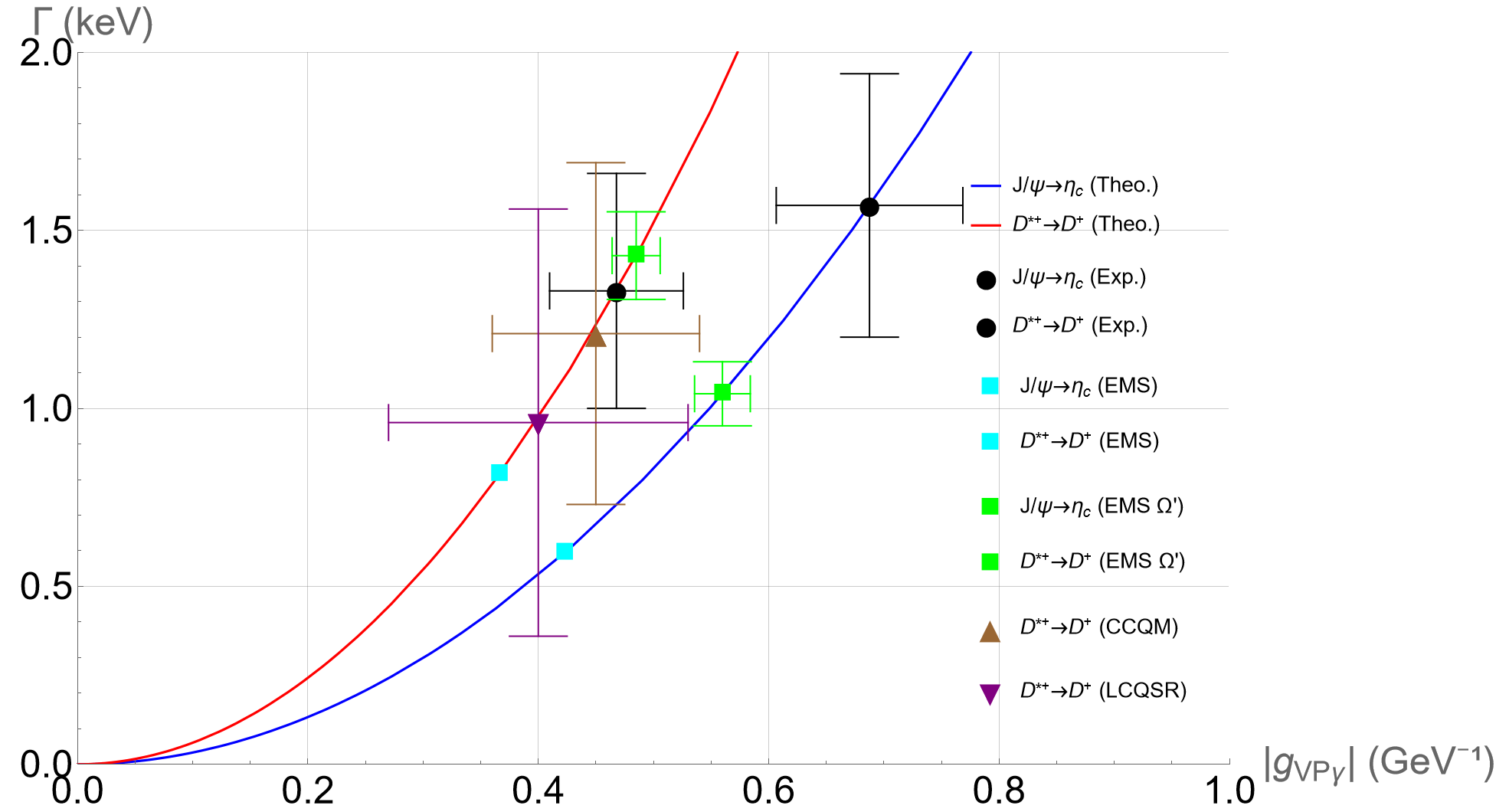}
    \caption{Variation of M1 decay width, $\Gamma$, with effective coupling, $|g_{VP\gamma}|$. Interestingly, the 2-loop anomalous magnetic moment corrections produce nearly identical $g_{VP\gamma}$ values, with error bars reflecting the uncertainties in $\Omega'$. }
    \label{Fig1}
\end{figure}

\begin{figure}[]
    \centering
    \includegraphics[width=0.8\linewidth]{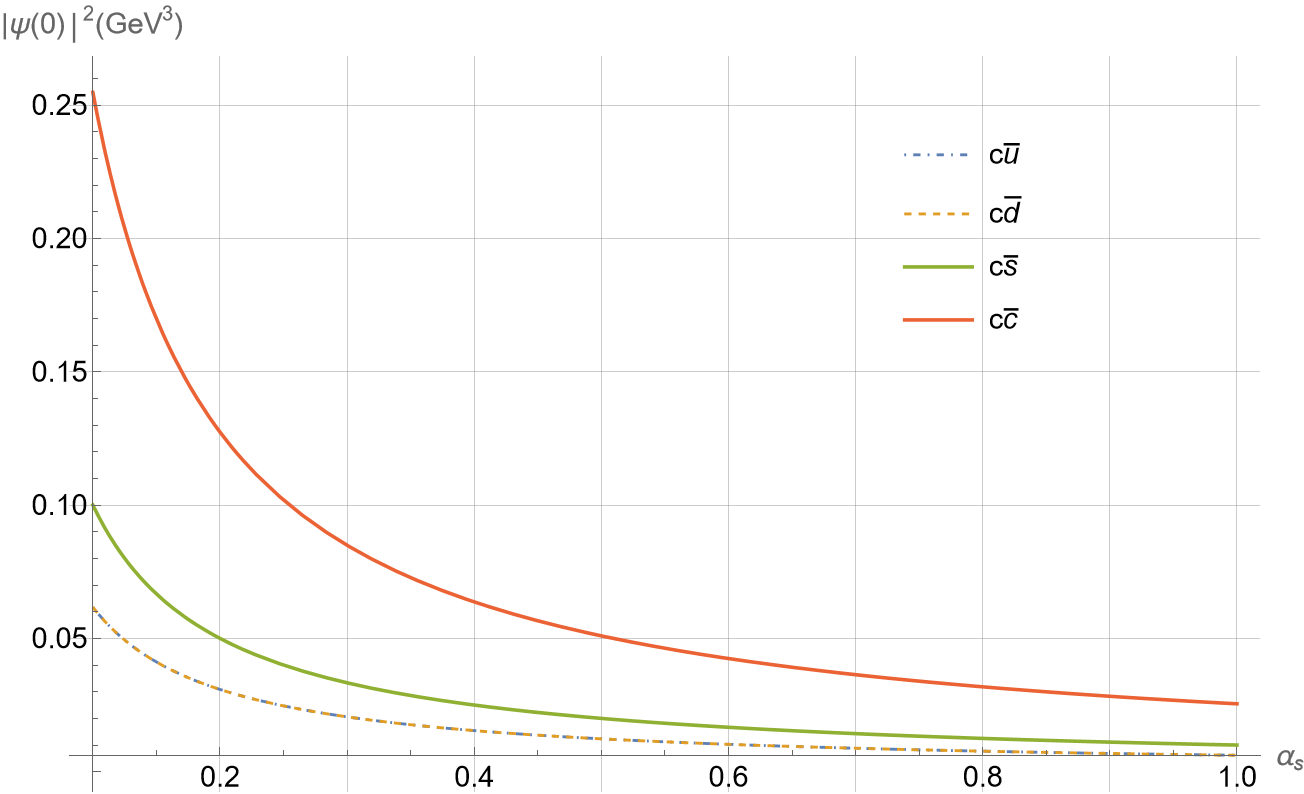}
    \caption{Variation of $|\psi(0)|^2$ estimates with $\alpha_s$ for charm mesons in EMS.}
    \label{Fig2}
\end{figure}

\begin{figure}[]
    \centering
    \includegraphics[width=0.8\linewidth]{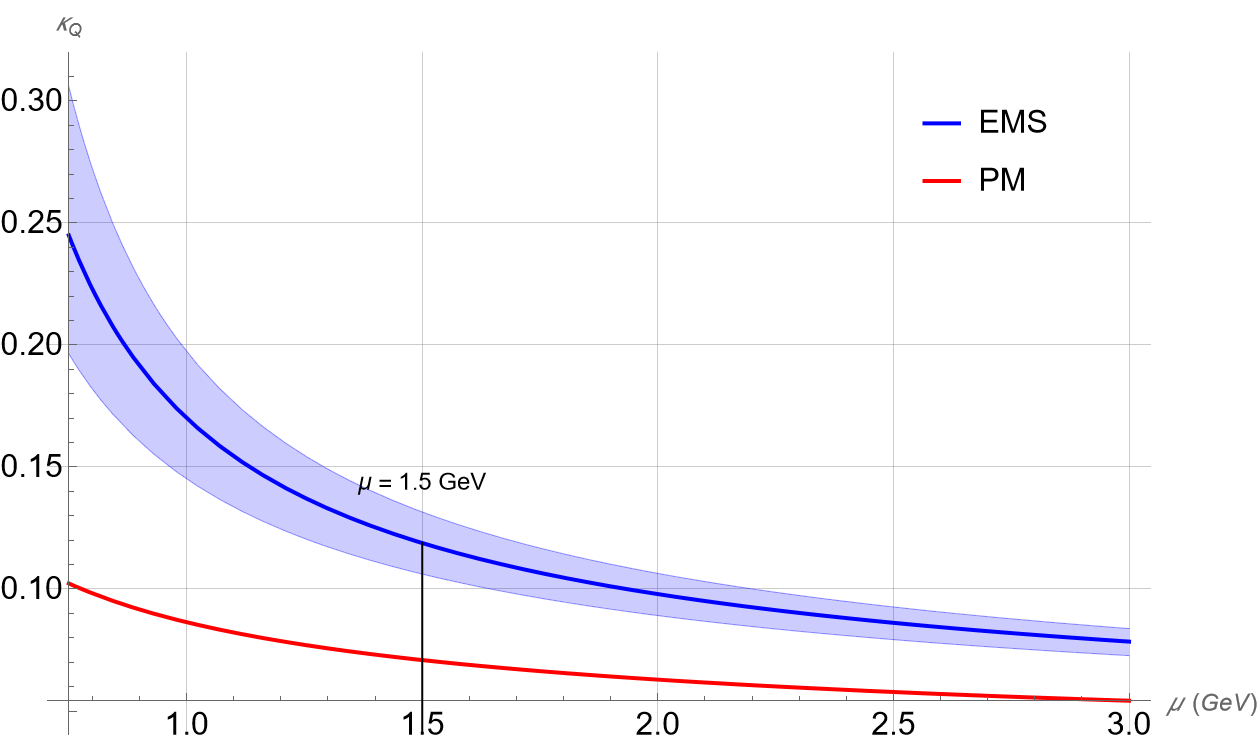}
    \caption{Variation of anomalous magnetic moment, $\kappa_Q = \frac{2\alpha_s(\mu)}{3\pi}$, with scale, $\mu$, for charm mesons. The corresponding curves in EMS and PM are plotted based on $\alpha_s(\mu)$ estimated from $\alpha_s(m_c)=0.56(6)$ and $\alpha_{sc}(m_c)=0.334$, respectively.}
    \label{Fig3}
\end{figure}

In this work, we predict the radiative M1 decay widths of ground state mesons involving $V \to P$ transitions up to charm sector in the EMS. We also calculate the magnetic and transition moments during the evaluation procedure utilizing the numerical inputs given in Table \ref{t1}. Though our focus is on charm meson transitions, we also present results for light mesons to refine our approach against existing experimental data. Our numerical results for magnetic moments of vector states up to charm mesons are listed in columns $2$ and $3$ of Table \ref{t3}, corresponding to EMS and CQM, respectively. Further, we calculate the transition moments and, consequently, the M1 decay widths using photon momenta given in Table \ref{t6}. We list our EMS predictions in column 2 of Tables \ref{t4}, \ref{t5}, \ref{t7}, and \ref{t8}. We wish to emphasize that, for reliable predictions, the accurate value of photon momenta plays a crucial role. Therefore, we use precise experimental masses of mesons available in PDG \cite{ParticleDataGroup:2024cfk} to reliably predict the aforementioned magnetic properties and photon momenta associated with the $V \to P$ transitions. Additionally, we employ a non-relativistic potential model to study these decays for charm mesons. The PM approach computes M1 decay matrix elements via spatial wave functions, whereas EMS derives them from spin-flavor wave functions through transition moments, providing a critical comparative test of both frameworks given the explicit wave function sensitivity of radiative M1 decays. We list our PM predictions for radiative M1 decay widths of charmed, charm-strange, and charmonium states in column $3$ of Table \ref{t8}, alongside EMS results. Our predictions are also compared with experimental values and those from other theoretical models. We further examine model sensitivities by studying the dependence of decay widths on the effective coupling for charm meson decays across various theoretical approaches, as illustrated in Fig.~\ref{Fig1}. We focus on the scale-dependence inherent to these processes, where the strong coupling and the wave function at the origin are intertwined. This interplay is quantified in Fig. \ref{Fig2}, which illustrates the sensitivity of $|\psi(0)|^2$ to $\alpha_s$ within the EMS for charm mesons. We report improved decay widths using phenomenological factors, with corresponding estimates of the wave function at the origin for charm mesons in both EMS and PM listed in Tables~\ref{t9} and \ref{t10}, respectively. Finally, the scale-dependence of the anomalous magnetic moment $\kappa_Q$ is analyzed for charm mesons in both EMS and PM (Fig. \ref{Fig3}). We discuss our numerical results in the following subsections. 

\subsection{Magnetic moments}
We present our magnetic moment results for the vector mesons up to charm sector in Table \ref{t3}. Our EMS results are obtained by utilizing the effective quark masses calculated from the inputs given in Table \ref{t1}. We also enlist CQM results obtained using our constituent quark mass inputs in order to understand the effect of strong hyperfine interaction contributions on the magnetic moments. The primary focus of our work is on charm mesons; however, we also present results for light flavor mesons to establish the reliability of our work by comparing them with existing experimental results. The following are our observations: 

\begin{enumerate}
\item It is interesting to note that precise experimental measurement of magnetic moment of mesons (in both light and heavy flavor sectors) is absent to date. The extremely short lifetimes of vector mesons, \textit{i.e.}, $\sim \mathcal{O}(10^{-23})$ s, pose difficulties to their magnetic moment measurements using standard techniques. However, an indirect estimate of the magnetic moment of $\rho^{+}$ meson is obtained using preliminary data from the BABAR Collaboration for the four-pion electroproduction process \cite{Druzhinin:2007cs}. Our EMS prediction for $\mu_{\rho^{+}}$ is in good agreement with this extracted result of $2.6(6)~\mu_N$ \cite{GarciaGudino:2015ocw} based on the BABAR data. Moreover, our result is also consistent with the predictions from various other theoretical models such as EBM \cite{Simonis:2018rld}, NJL \cite{Luan:2015goa}, DSE/BSE \cite{Xu:2024fun, Bhagwat:2006pu}, RH \cite{Badalian:2012ft}, and LQCD \cite{Owen:2015gva}. On the other hand, our CQM result yields a comparatively larger value, $3.068~\mu_N$ which is in good agreement with LCQSR \cite{Aliev:2009gj} prediction. However, the LCQSR \cite{Aliev:2009gj} result is compatible with other theoretical predictions only within their limits of error. Furthermore, it is also significant to point out that the experimental result has an uncertainty of $23\%$ and the predictions from all theoretical models lie within this uncertainty. Later, the BABAR Collaboration provided the cross-section measurement of $e^{+}e^{-}\to\pi^{+}\pi^{-}2\pi^{0}$ process with remarkable precision \cite{BaBar:2017zmc}. An updated result of $\mu_{\rho^{+}} = 3.3(4)~\mu_N$ with improved precision has been obtained very recently utilizing this data in a vector meson dominance approach \cite{Rojas:2024tmn}. Although the precision has been improved astonishingly, the central result has shifted to a higher numerical value that is currently outside the error bars of the previous estimate \cite{GarciaGudino:2015ocw} as well as most of the existing theoretical predictions. The difference in the treatment of model parameters unlike previous analysis \cite{GarciaGudino:2015ocw}, has resulted in a higher numerical value in Ref. \cite{Rojas:2024tmn}. However, it is interesting to note that the updated result for $\mu_{\rho^{+}}$ is comparable to our CQM prediction, though larger than the estimates of other theoretical works. New experimental observations from various collaborations can bring consensus on this matter.
		
\item Our EMS results for the magnetic moments of $K^{*+}$ and $K^{*0}$ are in good agreement with the predictions from other theoretical models \cite{Simonis:2018rld, Zhang:2021yul, Luan:2015goa, Aliev:2009gj, Bhagwat:2006pu, Xu:2024fun, Badalian:2012ft}, except LQCD \cite{Lee:2008qf}. Both LQCD and our CQM results yield slightly larger values than EMS and other approaches. It is worth to note that the magnetic moment of $K^{*0}$ is relatively small in magnitude with a negative sign due to the comparable but destructive contributions of $d$ and $s$ quark magnetic moments, with the former being the dominant component. Although the numerical value and sign of $\mu_{K^{*0}}$ is consistent in almost all theoretical models, LCQSR \cite{Aliev:2009gj} prediction is opposite in sign.  
		
\item In the charm sector, our EMS results for the magnetic moments of $D^{*0}$ and $D^{*+}$ are slightly larger than those from EBM \cite{Simonis:2018rld}, LCQSR \cite{Aliev:2019lsd}, MIT BM \cite{Zhang:2021yul}, NJL \cite{Luan:2015goa}, and DSE/BSE \cite{Xu:2024fun} predictions, but consistent with $\chi$PT \cite{Wang:2019mhm} predictions. It is worth to note that the magnetic moments of charm mesons are predominantly governed by the light quark magnetic moments as expected due to the smaller mass of light quark in the denominator of effective quark magnetic moment. The lighter quarks have larger magnetic moment contributions, \textit{i.e.}, $\mu_u^{\mathscr{E}}=1.933~\mu_N$, $\mu_d^{\mathscr{E}}=-0.967~\mu_N$, and $\mu_s^{\mathscr{E}}=-0.617~\mu_N$, compared to the charm quark magnetic moment, \textit{i.e.}, $\mu_c^{\mathscr{E}}=0.390~\mu_N$, due to the $1/m_q^{\mathscr{E}}$ dependence in the effective magnetic moment expression given by Eq. \eqref{eq12}. Interestingly, LCQSR \cite{Aliev:2019lsd} predicts substantially small numerical value for $\mu_{D^{*0}}$ as compared to all other theoretical models. In LCQSR calculations, the primary contribution to magnetic moments arises from the perturbative component of the spectral density. However, this contribution is negligible for neutral mesons with quarks and antiquarks of the same charge, leading to a small numerical value within the LCQSR framework, as noted in Ref. \cite{Aliev:2019lsd}. We wish to point out that the constructive contributions of $\mu_{\bar d}$ and $\mu_{\bar s}$ with $\mu_c$ lead to an overall magnetic moment with a positive sign in the case of $D^{*+}$ and $D_{s}^{*+}$ mesons. Furthermore, our numerical value of $\mu_{D_{s}^{*+}}$ is consistent with all the theoretical approaches excluding $\chi$PT \cite{Wang:2019mhm}, where it predicts a comparatively small value. A relatively large and unnatural $SU(3)$ symmetry breaking has been reported for this case in $\chi$PT \cite{Wang:2019mhm}. However, it should be emphasized that we have incorporated potential flavor symmetry breaking effects through effective quark masses in the current analysis. Moreover, $m_i$ and $b_{ij}$ terms in each flavor sector are calculated from the experimental masses of mesons in the respective sector. 
	
\item It is worth mentioning that, our EMS results are in agreement with CQM predictions in the case of charm mesons. This is followed by the reduction in the numerical value of strong hyperfine interaction terms as we go to the heavy sector. For charm mesons, the hyperfine contribution to magnetic moments is only about $4\%$, whereas in light mesons it is considerably larger, averaging around $20\%$. Furthermore, it is significant to highlight that neutral meson bound states containing quark flavors of same charge, \textit{i.e.}, $K^{*0}~\text{and}~D^{*0}$, exhibit negative sign in their magnetic moments. The negatively charged lighter quark (antiquark) flavor within a neutral meson contributes significantly to the overall magnetic moment, leading to the overall negative sign. However, LCQSR \cite{Aliev:2019lsd} predictions for the aforementioned neutral mesons contradict this observation, giving small positive numerical values for their magnetic moments. Therefore, the magnetic moment measurement of such particles can shed more light on the internal structure of mesons. 
\end{enumerate}
\subsection{Transition moments}
In this subsection, we present our results for the transition magnetic moments ($\mu_{V\to P\gamma}$) of light and charm mesons corresponding to $V\to P$ transitions, in Tables \ref{t4} and \ref{t5}, respectively. Our observations are listed below.

\begin{enumerate}
\item In the case of $\rho^{+(0)}\to\pi^{+(0)}$ transitions, our transition moment predictions are numerically equal due to the trivial effect of electromagnetic mass difference of pions on magnetic moments. It is worth to note that our transition moment results are larger in magnitude for $\omega\to\pi$ and $\rho\to\eta$ among other light meson transitions, due to the constructive contributions of $u$ and $d$ quark magnetic moments. On the other hand, $\omega\to\eta$ and $\phi\to\eta^{(\prime)}$ transition moments are numerically smaller and comparable. The $\omega\to\eta$ transition moment depends on the destructive contributions of $u$ and $d$ quark magnetic moments, while $\phi\to\eta^{(\prime)}$ transitions are governed by the strange quark magnetic moments. These results are also dependent on the $\eta-\eta^{\prime}$ mixing angle, $\phi_P$. It is interesting to note that the transition moments of $K^{*+}\to K^+$ and $K^{*0}\to K^0$ are numerically comparable but opposite in signs. The constructive contributions of $\mu_d$ and $\mu_s$ lead to an overall negative sign for $K^{*0}\to K^0$ transition moment, while the destructive contributions of $\mu_u$ and $\mu_s$ with a dominant contribution from $\mu_u$ result in a positive sign for $K^{*+}\to K^+$.

\item In the case of charm meson transitions, the magnetic moments of $u$ and $c$ quarks add constructively in the transition moment of $D^{*0}\to D^{0}$ leading to a larger value among other transitions. However, $D^{*+}\to D^{+}$ and $D_{s}^{*+}\to D_{s}^{+}$ transitions show smaller numerical values with negative sign for transition moments arising from the predominant contributions of negatively charged light quarks, \textit{i.e.}, $d$ and $s$, respectively, compared to the $c$ quark. It is interesting to note that $J/\psi \to \eta_{c}$ transition involves a non-zero transition moment, although the magnetic moment of the charmonium bound state is zero. This transition depends solely on the charm quark magnetic moment. As pointed out earlier, $J/\psi \to \eta_{c}$ transition moment is smaller in magnitude due to the $1/m_c^{\mathscr{E}}$ term that gives a smaller contribution. 

\item We observe that our EMS results for light meson transition moments are $1.5-2$ times larger than the EBM \cite{Simonis:2018rld, Simonis:2016pnh} and RPM \cite{Jena:2010zza} predictions. It should be noted that in our approach, constituent quark masses and $b_{ij}$ terms are calculated from experimental meson masses \cite{ParticleDataGroup:2024cfk}. However, some of the theoretical approaches, \textit{e.g.}, EBM \cite{Simonis:2018rld, Simonis:2016pnh}, $\chi$PT \cite{Wang:2019mhm}, adopt their inputs from the experimental masses and magnetic moments of baryons. This results in a considerable difference in the predictions from these works, which is more profound in the light quark sector. Specifically, our constituent quark masses are smaller than those of other theoretical models, which results in our transition magnetic moments being larger than other works. In addition to this, EBM \cite{Simonis:2018rld, Simonis:2016pnh} and RPM \cite{Jena:2010zza} explicitly incorporate scale factors into their calculation of transition moments. The EBM \cite{Simonis:2018rld, Simonis:2016pnh} has implemented the use of distinct scale factors for light and heavy quarks that account for the effects of center-of-mass motion, recoil, and other corrections on the magnetic observables. Furthermore, these scale factors are usually optimized to reproduce the experimental decay widths of light and heavy mesons, specifically, $\rho\to\pi\gamma$ and $\omega\to\pi\gamma$ for the light mesons, and $J/\psi\to\eta_{c}\gamma$ for the heavy mesons \cite{Simonis:2018rld}. On the other hand, in RPM \cite{Jena:2010zza}, the recoil of the daughter meson is incorporated as a scale factor in the transition moment calculations to account for the effective momentum transfer involved in the decays. This results in numerical differences for transition moments in comparison with ours. In particular, the definition and numerical value of transition moments differ between our EMS approach and other theoretical models; however, the corresponding M1 decay widths are expected to agree.
	
\item As observed in the case of light mesons, our EMS results corresponding to $D^{*+}\to D^{+}$ and $D^{*0}\to D^{0}$ are approximately 1.5 times larger in comparison with EBM \cite{Simonis:2018rld}, $\chi$PT \cite{Wang:2019mhm} and RPM \cite{Jena:2002is} predictions. However, our results are consistent with the predictions of CQM \cite{Wang:2023bek} due to comparable numerical values of constituent quark masses. It is worth to note that for $D^{*+}_{s}\to D^{+}_{s}$ and $J/\psi\to\eta_{c}$ transitions, our results are in good agreement with the aforementioned theoretical approaches. Thus, it is evident that the agreement among different theoretical models improves as we move to the heavier mesons, indicating the minimal effect of scale factor in the heavy sector. Interestingly, the phase space factor becomes unity for radiative decays involving heavy mesons due to the comparable masses of parent and daughter particles for the approaches \cite{Simonis:2018rld, Jena:2002is}. This indicates the requirement of scale factors in $D^*\to D$ transitions to explain the experimental results. As discussed earlier, the smaller mass of lighter quarks leads to larger contributions of light quark magnetic moments in the transition moments. Therefore, the heavy quark contributions are expected to be suppressed leading to predominant contributions by light quarks. Similar observations are reported in LFQM \cite{Choi:2007se}, and RIQ \cite{Priyadarsini:2016tiu} in which the dynamical treatment of transition moments through momentum-dependent transition form factors is employed. For $D^{*+}_{s}\to D^{+}_{s}\gamma$ and $J/\psi\to\eta_{c}\gamma$ decays, the recoil effects of final state pseudoscalar mesons are negligible indicating minimal momentum-dependence of transition form factor in these decays. Consequently, the radiative M1 decay width predictions in the charm sector are expected to be comparable among different theoretical models including ours.
		
\item It is crucial to note that the choice of quark transition mass significantly impacts the calculation of light meson transition moments; and subsequently their M1 decay widths. Employing the arithmetic mean of effective quark masses as the transition mass over the geometric mean results in a decrease in transition moments. It may be noted that the arithmetic mean offers equal distribution of effective mass, while the geometric mean leads to the lower effective mass value. However, this reduction is less significant in the case of heavy mesons, because the arithmetic and geometric means of effective quark masses become numerically comparable, as we move from light sector to heavy sector. Given the lack of experimental data on transition moments, this approach appears to be a reasonable choice for explaining the decay widths. We find that using the arithmetic mean as the transition mass yields excellent results for M1 decay widths compared to the geometric mean, especially for light mesons. 
\end{enumerate}
\subsection{M1 decay widths}
We present our EMS results for the radiative M1 decay widths of light and charm mesons in Tables \ref{t7} and \ref{t8}, respectively. We also list our predictions obtained from PM using Cornell potential for charm mesons in column 3 of Table \ref{t8}. In addition, we compare our results with the experimental data \cite{ParticleDataGroup:2024cfk} as well as predictions from various theoretical models. The key findings based on the observations with respect to EMS and PM are as follows.
\subsubsection{\textbf{EMS framework}}
\label{EMS_m1}
\begin{enumerate}		
\item The theoretical predictions for M1 decay widths are expected to follow a trend similar to those of transition magnetic moments, given that the widths depend on both the magnitude of the transition moment and kinematic factors. For light meson radiative transitions, our decay width results for $\rho^{+}\to\pi^{+}\gamma$ and $\rho^{0}\to\pi^{0}\gamma$ are in very good agreement with the experimental values \cite{ParticleDataGroup:2024cfk}, while our prediction for $\Gamma_{\omega\to\eta\gamma}$ is consistent with experimental data within the error bars. It should be noted that the experimental data on M1 decay widths of light mesons involve uncertainties greater than $10\%$ and require new direct measurements with improved precision. Interestingly, M1 partial decay widths extracted from measurements of branching ratio and total width of the decaying particle have less uncertainties than the available direct measurements. Therefore, we have listed these extracted values of M1 decay widths under the label ``PDG" in Table \ref{t7} and compared our results with them. We observe that our results for $\rho\to\eta\gamma$, $\omega\to\pi\gamma$, and $\omega\to\eta\gamma$ are consistent with these extracted results of M1 decay widths \cite{ParticleDataGroup:2024cfk}. The $\omega\to\pi\gamma$ decay width is the largest among all transitions due to the larger values of transition moment and photon momentum. On the other hand, our results for $\Gamma_{K^{*+}\to K^{+}\gamma}$ and $\Gamma_{K^{*0}\to K^{0}\gamma}$ show roughly $30\%$ deviation from the experimental data. Furthermore, our predictions for $\phi\to\eta\gamma$ and $\phi\to\eta^{\prime}\gamma$ are smaller than the experimental values by roughly a factor of $2$. The photon momentum is very small for $\phi\to\eta^{\prime}$ transition, which leads to a decay width smaller than any other light meson transition. 
		
\item A noteworthy trend is the reduction in decay width as one transitions from lighter to heavier meson sectors. Consistent with observations in transition moments, the M1 decay width for $D^{*0}\to D^{0}\gamma$ is the largest among charm meson $V\to P\gamma$ decays. While experimental measurements of charm meson M1 widths are currently unavailable, Table \ref{t8} presents experimental decay widths for $D^{*+}\to D^{+}\gamma$ and $J/\psi\to \eta_{c}\gamma$, sourced from PDG data \cite{ParticleDataGroup:2024cfk} based on branching ratios and total decay widths. Our EMS predictions are smaller than these values by $39\%$ and $62\%$, respectively. Considering the inherent experimental uncertainties, typically around $25\%$, we look forward to new experimental data offering improved precision.

\item Our theoretical predictions are systematically compared with existing literature for both light and heavy-flavored meson systems. In the light meson sector, our M1 radiative decay widths demonstrate excellent agreement with theoretical frameworks such as EBM \cite{Simonis:2018rld, Simonis:2016pnh}, RPM \cite{Jena:2010zza}, LFHM \cite{Ahmady:2020mht}, LFQM \cite{Choi:1997iq}, and PM(AP1) \cite{Bonnaz:2001aj}. A key advantage of EMS is its ability to yield consistent results without requiring a scale factor. This stems from the distinct nature of one-gluon exchange interaction between quark-(anti)quark pairs in mesons compared to baryons, particularly in the light quark sector. The difference in this dynamics between these systems leads to varying effective interactions. However, in the heavy quark sector, this distinction diminishes due to the reduced hyperfine interaction contributions, yielding nearly equivalent effective quark masses for mesons and baryons. This stands in contrast to other theoretical approaches, which commonly rely on such empirical scale factors to align their predictions with experimental measurements. Consequently, while our calculated transition magnetic moments may differ numerically from other theoretical predictions, the resulting decay widths are in excellent agreement with experimental measurements, with few exceptions. Specifically, other models overestimate radiative decay widths for processes involving $\eta^{(\prime)}$ mesons in the final state by as much as $(30-90)\%$ relative to experimental data. The EMS approach demonstrates particular robustness in reproducing experimental decay widths for challenging transitions such as $\omega \to \eta\gamma$ and $K^{*+}\to K^{+}\gamma$, where other theoretical frameworks show significant discrepancies. For instance, the PM(AP1) model \cite{Bonnaz:2001aj} predicts a decay width for $K^{*+}\to K^{+}\gamma$ that exceeds the experimental value by a factor of $\sim 2$, while NJL-based calculations \cite{Deng:2013uca} underestimate widths by factors of $2$–$3$ due to the absence of quark confinement mechanism in the model. Within NJL, constituent quark masses are phenomenologically adjusted to larger values to ensure meson binding energies remain below the constituent quark-antiquark threshold, thus maintaining bound-state stability in the absence of explicit confinement dynamics. Similar discrepancies of approximately a factor of two are observed in DSE calculations \cite{Maris:2002mz}, particularly for the $\omega\to\pi\gamma$ radiative decay width.

\item In the heavy-flavored (charm) meson sector, our predictions are comparable to the results from diverse theoretical frameworks, including the EBM \cite{Simonis:2018rld}, LCQSR \cite{Pullin:2021ebn}, NJL \cite{Deng:2013uca}, LQCD \cite{Becirevic:2009xp}, $\chi$PT \cite{Wang:2019mhm}, LFQM \cite{Choi:2007se}, RIQ \cite{Priyadarsini:2016tiu}, RQM \cite{Ebert:2002xz}, and CM \cite{Cheung:2014cka}, with notable exceptions in specific cases. Theoretical approaches to M1 decay widths can be broadly categorized into two main methodologies. The first relies on constituent quark models (and potential models) to elucidate meson spectroscopy and decay properties through calculations of static magnetic moments. The second employs more intricate theoretical frameworks, such as LFQM, RIQ, and QCDSR, to investigate decay dynamics by obtaining $g_{VP\gamma}$ couplings (electromagnetic form factors) for theoretical decay widths. It is noteworthy that most theoretical models report discrepancies in the decay width results for the $D^{*+}\to D^+\gamma$ decay as compared to the experiment. However, these theoretical predictions come with significant uncertainties. These uncertainties stem from factors like quark masses and higher-order contributions, such as $\mathcal{O}(p^4)$ in $\chi$PT \cite{Wang:2019mhm}, as well as large statistical errors in LQCD \cite{Becirevic:2009xp}. Among other theoretical results, the predictions from PM(AP1) \cite{Bonnaz:2001aj} and GI \cite{Godfrey:2015dva} are considerably larger than both our findings and the experimental data, while QCDSR \cite{Lu:2024tgy} predictions are several times smaller. For the $J/\psi\to\eta_{c}\gamma$ M1 decay width, results from EBM \cite{Simonis:2018rld}, LFQM \cite{Choi:2007se}, PM(AP1) \cite{Bonnaz:2001aj}, and more recently, self-consistent LFQM \cite{Ridwan:2024ngc} are consistent with the current experimental value. It is important to note that EBM scaled their charm sector results using experimental decay width of $J/\psi\to\eta_{c}\gamma$. In contrast, decay width predictions from LQCD \cite{Colquhoun:2023zbc}, RIQ \cite{Priyadarsini:2016tiu}, and RQM \cite{Ebert:2002pp} are significantly larger compared to both our EMS result and the experimental value.

\item In general, the discrepancies between theoretical predictions and experimental values remain a persistent challenge in studies of radiative M1 decay widths for $V \to P$ meson transitions. As stated previously, while most models employ a scale factor to reconcile light meson decay predictions with experimental data, EMS achieves agreement without any scale adjustment. Furthermore, theoretical predictions for heavy meson decay widths exhibit significant discrepancies with experimental observations across all models, prompting the use of empirical scale factors in certain frameworks. Notably, all existing theoretical approaches, including those based on electromagnetic form factors or effective couplings, failed to yield consistent predictions, with few exceptions. This highlights the complexity of the underlying decay dynamics for M1 decays. However, the differences between these approaches may provide a way to estimate an effective scale factor. A key observation is the direct proportionality between the effective coupling $g_{VP\gamma}$ and the static transition magnetic moment $\mu_{V \to P} $, which emerges when comparing M1 decay width expressions across different frameworks. Specifically, the relation
\[  
g_{VP\gamma} \propto \mu_{V \to P},  
\]  
 holds independently of additional mass-dependent factors, such that $\mu_{V\to P}$ expressed in units of nuclear magneton, $\mu_N$. This suggests that $ \mu_{V \to P} $ could serve as a fundamental input for determining the scale factor in a model-independent manner. Recent theoretical calculations of the $D^{*+}\to D^{+}\gamma$ coupling, $g_{VP\gamma}$, yield values of $-0.45(9) \text{ GeV}^{-1}$ from CCQM \cite{Tran:2023hrn} and $0.40^{+0.12}_{-0.13} \text{ GeV}^{-1}$ from LCQSR \cite{Pullin:2021ebn}. Experimentally, this coupling is determined to be $-0.47(7) \text{ GeV}^{-1}$ \cite{ParticleDataGroup:2024cfk}. An averaged theoretical value of $g_{VP\gamma}$ across various models \cite{Tran:2023hrn} is $-0.35(15) \text{ GeV}^{-1}$. Our calculated coupling, $g^{EMS}_{VP\gamma} = -0.366 \text{ GeV}^{-1}$, derived from our transition magnetic moment using the aforementioned relation, shows reasonable agreement with this theoretical average. However, experimental data suggest a numerically larger value for the coupling. Consequently, a scale factor, $\Omega$, obtained through a comparison of static and dynamical transition magnetic moments, is necessary to reconcile our results with experimental observations. Thus, considering $g^{Expt}_{VP\gamma} = -0.47(7) \text{ GeV}{}^{-1}$ \cite{ParticleDataGroup:2024cfk} and our estimate for $D^{*+}\to D^{+}\gamma$, we obtain 
\[ \Omega = \frac{g^{Expt}_{VP\gamma}}{g^{EMS}_{VP\gamma}} = \frac{\mu^{Expt}_{V\to P}}{\mu^{EMS}_{V\to P}} = 1.28.\]
This factor will incorporate dynamical effects into our static transition magnetic moments, and thereby improve our transition moments and M1 decay widths of charm mesons\footnote{Note that $\Omega$ could also be extracted in a similar fashion using theoretical information \cite{Tran:2023hrn} and was found to be $1.23$, which is comparable to the scale factor calculated from experimental $g_{VP\gamma}$ \cite{ParticleDataGroup:2024cfk}.}. Also, numerical values of $g_{VP\gamma}$ for $D^{*+}\to D^{+}\gamma$ from the experiment \cite{ParticleDataGroup:2024cfk} and CCQM \cite{Tran:2023hrn} contain uncertainties $\sim 14\%$ and $\sim 20\%$, respectively. Therefore, the estimation of the scale factor may vary with improvements in experimental measurements. A comparison of $g_{VP\gamma}$ across different approaches, as illustrated in Fig.~\ref{Fig1}, shows that although theoretical estimates from CCQM \cite{Tran:2023hrn}, LCQSR \cite{Pullin:2021ebn}, VMD-inspired methods, and our EMS determination differ numerically, they consistently predict values of the same sign and order of magnitude, overlapping within the quoted $\sim (10-20)\%$ uncertainties. The experimental determination for $D^{*+}\!\to D^{+}\gamma$, $g^{ Expt}_{VP\gamma}=-0.47(7)\,\text{GeV}^{-1}$ \cite{ParticleDataGroup:2024cfk}, exceeds our static EMS estimate $g^{ EMS}_{VP\gamma}=-0.366\,\text{GeV}^{-1}$ by a factor $\Omega\simeq 1.28$, indicating a $\sim\!28\%$ dynamical enhancement beyond the static treatment. It may be emphasized that VMD-based approaches offer a phenomenologically transparent framework that correctly reproduces the scale and sign of the coupling. However, they remain sensitive to off-shell effects, form factor choices, and input parameters such as quark masses. This supports the conclusion that, while long-distance dynamics are reasonably well described, short-distance contributions and relativistic corrections must be effectively absorbed into the scale factor $\Omega$ to achieve quantitative agreement with experiment. Given the lower uncertainties of $\Gamma(D^{*+}\to D^{+}\gamma)$, we chose $\Omega = 1.28$ to refine our transition magnetic moments uniformly throughout the charm sector, beginning with $D^{*+} \to D^{+}$ transition. Thereby, our M1 decay widths, which were previously underestimated, are vastly improved, and are listed in column 2 of Table \ref{t9}. We observe that the deviation of $\Gamma_{J/\psi\to \eta_{c}\gamma}$ with respect to the experimental value has decreased from $62\%$ to $38\%$. In addition, our improved EMS results for $\Gamma_{D^{*0}\to D^{0}\gamma}$ and $\Gamma_{D^{*+}_{s}\to D^{+}_{s}\gamma}$ are consistent with theoretical predictions from EBM \cite{Simonis:2018rld}, $\chi$PT \cite{Wang:2019mhm}, LFQM \cite{Choi:2007se}, RIQ \cite{Priyadarsini:2016tiu}, and RQM \cite{Ebert:2002xz}. Interestingly, our improved prediction for $\Gamma_{D^{*0}\to D^{0}\gamma}$ is in excellent agreement with the recent theoretical estimate of $19.4$ keV based on relativistic corrections for the instantaneous BS approach \cite{Jia:2024imm}. Furthermore,  J. Wang \textit{et al.} \cite{Wang:2025fzj} has used effective Lagrangian approach to provide a reasonable estimate for $\Gamma_{D^{*+}_{s}\to D^{+}_{s}\gamma}=0.161(16)$ keV utilizing  recent BESIII \cite{BESIII:2022kbd} measurements, that shows good agreement with EMS predictions with and without scale factor. Although introducing a scale factor improves EMS predictions for charm meson decays, discrepancies with experiment persist due to the intrinsic scale-dependence of QCD, where $\alpha_s$ and $|\psi(0)|^2$ are intertwined. The factor $\Omega$ can thus be interpreted as effectively incorporating short-distance QCD effects beyond the static quark model. A detailed discussion of these scale-dependent corrections is provided in Sec.~\ref{scale}.

\item We note that the M1 decay widths for the transitions $ D^{*0} \to D^{0}\gamma $ and $ D^{*+}_{s} \to D^{+}_{s}\gamma $ have not yet been experimentally measured. Furthermore, existing theoretical predictions for $D^{*0} \to D^{0}\gamma $ exhibit significant variations (as shown in Table \ref{t8}), with reported widths ranging from a few keV to approximately $100$ keV, spanning two orders of magnitude. Interestingly, the average theoretical value of the coupling $ g_{VP\gamma} $ for this decay, $ 1.44(25) \text{ GeV}^{-1} $ \cite{Tran:2023hrn}, also reflects a larger variation. Given the absence of experimental data and the substantial theoretical uncertainty, the $ D^{*0} \to D^{0}\gamma $ transition is less suitable for reliably extracting the scale factor at this time. Similarly, we observe that the decay width prediction of $D^{*+}_{s}\to D^{+}_{s}\gamma$ is considerably small in NJL \cite{Deng:2013uca}, LQCD \cite{Meng:2024gpd}, and QCDSR \cite{Lu:2024tgy}, while exceptionally large in LCQSR \cite{Pullin:2021ebn}, CM \cite{Cheung:2015rya}, and GI \cite{Godfrey:2015dva} compared to other theoretical models by an order of magnitude. Therefore, precise experimental measurements of these decay channels are essential to resolve the disagreements among various theoretical predictions.

\item Furthermore, for $D^{*0}\to D^{0}$ and $D^{*+}_{s}\to D^{+}_{s}$ transitions, only the upper limit of the total widths of $D^{*0}$ and $D^{*+}_{s}$ mesons are available. Thus, the decay widths of $D^{*0}\to D^{0}\gamma$ and $D^{*+}_{s}\to D^{+}_{s}\gamma$ decays are found to have an upper limit of $741.3(18.9)$ keV and $1776.5(13.3)$ keV, respectively, which are significantly larger than all theoretical predictions, including ours. Therefore, experiments need improvement in their sensitivity for the precise determination of the aforementioned decay widths. Nevertheless, we try to estimate the total width of $D^{*0}$ using our calculated M1 decay widths and experimental branching ratios (given in PDG \cite{ParticleDataGroup:2024cfk}) through the following relation:
\begin{equation*}
    \Gamma_{tot}(D^{*0}) = \Big(\frac{BR^{expt}(D^{*+}\to D^{+}\gamma)}{BR^{expt}(D^{*0}\to D^{0}\gamma)}\Big)\Big(\frac{\Gamma^{EMS}(D^{*0}\to D^{0}\gamma)}{\Gamma^{EMS}(D^{*+}\to D^{+}\gamma)}\Big)\Gamma^{expt}_{tot}(D^{*+}).
\end{equation*}
In a similar way, the total width of $D_s^{*+}$ can also be calculated by replacing the branching ratio and M1 decay width of $D^{*0}$ by $D_s^{*+}$ in the above relation. Thus, we obtained the total widths of $D^{*0}$ and $D_s^{*+}$ as $\Gamma_{tot}(D^{*0}) = 54.41$ keV and $\Gamma_{tot}(D_s^{*+}) = 0.23$ keV from EMS. Our EMS results are in very good agreement with $\Gamma_{tot}(D^{*0}) = 55(6)$ keV and $\Gamma_{tot}(D_s^{*+}) = 0.19(1) \text{ keV} ~\{0.172(16) \text{ keV}\} $, reported in LFQM \cite{Choi:2007se} $\{$effective Lagrangian approach \cite{Wang:2025fzj}$\}$, and $\Gamma_{tot}(D^{*0}) = (53\pm5\pm7) \text{ keV} ~\{54 \text{ keV}\}$, stated in LQCD \cite{Becirevic:2012pf} $\{$instantaneous BS approach \cite{Jia:2024imm}$\}$.
\end{enumerate}

\subsubsection{\textbf{PM approach}}
\label{PM_m1}

In the potential model approach, we compute the isoplet masses (and wave functions) of the bound states by finely optimized parameters (Table \ref{tab:2_Potentialparameters}). We obtain\footnote{Note that since we have chosen $m_u = m_d = 0.325$ GeV (Table \ref{tab:2_Potentialparameters}), we do not differentiate between $D^0$ and $D^+$ in the case of their bound state masses.}
\[
M_{D} = 1870.20 \text{ MeV,} \qquad M_{D^*} = 2010.96 \text{ MeV,} 
\]
\[
M_{D_s} = 1968.80 \text{ MeV,} \qquad M_{D_s^*} = 2112.71 \text{ MeV,}
\]
\[
M_{\eta_c} = 2983.74 \text{ MeV, and } M_{J/\psi} = 3096.68 \text{ MeV;}
\]
which deviate less than $1\%$ from their experimental values \cite{ParticleDataGroup:2024cfk}. This ensures that the photon momenta ($k_\gamma$) calculated from our theoretical masses is also in good agreement with the experimental results, as shown in the column $3$ of Table \ref{t6}. We use the estimated wave function of the bound states, along with $k_\gamma$, to predict the M1 decay widths of corresponding meson transitions. We observe the following:

\begin{enumerate}
\item  We list our M1 decay widths of charm mesons, computed using PM approach, in column $3$ of Table \ref{t8}. Our predictions from the PM approach follow a similar trend as observed in EMS, we found that $D^{*0}\to D^{0}\gamma$ has the largest M1 width, while $D^{*+}_{s}\to D^{+}_{s}\gamma$ is the smallest. However, the observed M1 decay widths are substantially larger than the EMS. The smaller M1 widths involving charged meson states can be attributed mainly to the second term in the matrix element of Eq. \eqref{eq21}, which provides the dominant contribution. Specifically, for the charmed and charm-strange mesons, the charge and mass of the light antiquark ($\frac{e_{\overline{q}^\prime}}{m_{\overline{q}^\prime}}$), combined with the negative sign in second term, lead to a smaller overall magnitude for the matrix element of the charged $D^{*+}$ and $D^{*+}_{s}$ states. Furthermore, the larger mass of $m_{\overline{s}}$ leads to a much lower value of the matrix element of $D^{*+}_{s}$, resulting in the smallest M1 decay width among the charm sector states. Conversely, for the $D^{*0}$ state, the larger charge and the smaller mass of the $u$ quark result in a larger magnitude of $\frac{e_{\overline{u}}}{m_{\overline{u}}}$, which contributes to the larger magnitude of the matrix element, thereby, larger M1 decay width. Similarly, the overall magnitude of the $J/\psi\to \eta_{c} \gamma$ matrix element is larger than those of the charged states, but smaller than $D^{*0}\to D^{0}\gamma$, resulting in the second largest M1 width among the charm sector mesons. Comparing with experimental results, we find that our predictions are relatively larger by $34\%$ and $66\%$ for the $D^{*+}\to D^{+} \gamma$ and $J/\psi\to \eta_{c} \gamma$ decays, respectively. In the case of $D^{*0}\to D^{0}\gamma$ and $D^{*+}_{s}\to D^{+}_{s}\gamma$ decays, our predictions are well within the experimental upper bounds \cite{ParticleDataGroup:2024cfk}.

\item We emphasize that the $k_\gamma$, in Eq. \eqref{eq21}, though small in charm mesons, is crucial for the accurate estimation of M1 decay widths as it appears as $k_\gamma^3$ and is also an argument of the spherical Bessel function. Given the importance of precise determination of these quantities, the discrepancy between the calculated M1 decay widths and experimental values (observed in $D^{*+}\to D^{+} \gamma$ and $J/\psi\to \eta_{c} \gamma$ decays) suggests a potential overestimation of the matrix element. Interestingly, $\alpha_s$ which played a vital role in the accurate determination of low-lying state masses, shows a negligible direct effect on the M1 decay widths. Similar behavior is also observed for the other potential model parameters. However, it should be noted that their influence (particularly from $\alpha_s$) on the bound state wave function persists \cite{A:2023bxv, Godfrey:1985xj}, which in turn affects the M1 widths. Additionally, we noticed that the mass of the quark has considerable impact on the decay width; with the decrease in $m_q$ causing an increase in the decay width values, as expected from Eq. \eqref{eq21}. A similar observation can be made from our EMS approach as well. Thus, the overestimation of the matrix element and the $\alpha_s$ dependence of the wave function suggests that further improvement to the prediction of M1 decay widths could only be brought about from the wave function of the bound states.

\item Comparing our PM width for $J/\psi\to \eta_{c} \gamma$ decay with other theoretical works, we observe that our predictions are in general larger than other works, however, are compatible with LQCD \cite{Colquhoun:2023zbc} and RIQ \cite{Priyadarsini:2016tiu}. Similarly, for the $D^{*+}\to D^{+}\gamma$ decay, other approaches quote smaller M1 decay width values than ours, except for potential model-based approaches such as PM(AP1) \cite{Bonnaz:2001aj} and GI \cite{Godfrey:2015dva}. Furthermore, our M1 decay widths from PM for $D^{*0}\to D^{0}\gamma$ and $D^{*+}_{s}\to D^{+}_{s}\gamma$ decays are larger than EMS by a factor of $\sim 3$ and $\sim 2$, respectively. However, our $D^{*0}\to D^{0} \gamma$ width is in good agreement with LQCD \cite{Becirevic:2009xp}, while PM(AP1) \cite{Bonnaz:2001aj} and GI \cite{Godfrey:2015dva} quote larger numerical values; the results of the other theoretical approaches are smaller than ours. Significant variation is observed among the models for the $D^{*+}_{s}\to D^{+}_{s}$ M1 transition, with QCDSR \cite{Lu:2024tgy} and CM \cite{Cheung:2015rya} quoting the smallest and largest numerical value, respectively. Nevertheless, our prediction is in agreement with $\chi$PT \cite{Wang:2019mhm} and PM(AP1) \cite{Bonnaz:2001aj}. Furthermore, we estimated the total widths of $D^{*0}$ and $D^{*+}_{s}$ using PM as $\Gamma_{tot}(D^{*0}) = 75.19$ keV and $\Gamma_{tot}(D_s^{*+}) = 0.23$ keV, respectively. Our PM value for $\Gamma_{tot}(D_s^{*+})$ is in good agreement with results from EMS ($\Gamma_{tot}(D_s^{*+}) = 0.23$ keV), LFQM ($\Gamma_{tot}(D_s^{*+}) = 0.19(1)$ keV) \cite{Choi:2007se}, and effective Lagrangian approach ($\Gamma_{tot}(D_s^{*+}) = 0.172(16)$ keV) \cite{Wang:2025fzj}. Similarly, our $\Gamma_{tot}(D^{*0})$ agrees well with the $\chi$PT estimate ($\Gamma_{tot}(D^{*0}) = 77.7^{+26.7}_{-20.5}$ keV) \cite{Wang:2019mhm}. This lack of consensus among the models highlights the challenges in achieving consistent predictions that align well with experimental values.

\item Aforementioned, it is possible to improve our PM results for the M1 decay widths of charm mesons through their bound state wave functions which govern the matrix element. The sensitivity of the wave function to phenomenological parameters ($\alpha_s$ and $\sigma$) implies that parameters optimized for mass predictions also influence the wave function itself. This suggests that the scale-dependence of the parameter $\alpha_s$ could lead to a scale-dependence in the wave function \cite{A:2023bxv}. Furthermore, this can also be interpreted as $\alpha_s$ having a flavor dependence due to the constituents of the bound state, which is reflected in the wave function. To address this, we introduce a phenomenological correction factor, $f$, to account for this ambiguity in the scale and flavor dependence of the wave function\footnote{Such a correction factor is not dissimilar to the scale factor ($\Omega$) we have incorporated to improve our EMS predictions (as discussed previously).}. Moreover, as inferred from Ref. \cite{Godfrey:1985xj}, the leading behavior of matrix elements can be modified through factors fitted from experiment. In other words, experimental data can be used to refine the matrix elements, therefore, we use the flavor-symmetrical charmonium system to extract this scale factor, specifically from the experimental M1 decay width of the $J/\psi$. Thus, using $\Gamma^{Expt}(J/\psi \to \eta_{c}\gamma) = 1.57(37)$ keV \cite{ParticleDataGroup:2024cfk} and our theoretical value (from Eq. \eqref{eq21}), we find
\[
f = \frac{\Gamma^{Expt}(J/\psi \to \eta_{c}\gamma)}{\Gamma^{PM}(J/\psi \to \eta_{c}\gamma)} = 0.603.
\]
Subsequently, this factor, which represents a $40\%$ reduction, is applied to all our predictions for the decay widths of charmed and charm-strange mesons. The refined predictions, listed in column $4$ of Table \ref{t9}, show notable improvement. Specifically, our predicted M1 decay width for $D^{*+}\to D^{+}\gamma$ decay is now in good agreement with the experimental value within uncertainty. Furthermore, our improved M1 predictions are consistent with other theoretical works, with few exceptions. Our improved $\Gamma_{D^{*0}\to D^{0}\gamma}$ prediction is in good agreement with the $19.4$ keV theoretical estimate from the relativistic BS approach \cite{Jia:2024imm}. Similarly, our PM approach prediction for $\Gamma_{D^{*+}_{s}\to D^{+}_{s}\gamma}$ show excellent agreement with the $0.161(16)$ keV estimate derived from BESIII data \cite{BESIII:2022kbd} in the effective Lagrangian approach \cite{Wang:2025fzj}. Finally, including this correction factor significantly reduces the disagreement between our EMS and PM predictions. We emphasize that, although the phenomenological factor $f$ improves agreement with experimental data, its necessity also indicates the presence of missing corrections in the current approach to M1 decay widths. In particular, the scale-dependence of $\alpha_s$ and modifications to the wave function are expected to introduce non-negligible corrections to the M1 matrix element. Consequently, the factor $f$ may be interpreted as effectively incorporating QCD effects, such as the anomalous magnetic moment and relativistic corrections, that lie beyond the scope of a static potential model. A more detailed discussion of this interpretation is presented in the following section.
\end{enumerate}

\subsection{Scale-dependent anomalies in M1 decay widths}
\label{scale}
\subsubsection{Scale-dependence of wave function}

The calculation of M1 transition widths is constrained by the QCD scale-dependence, which intertwines the strong coupling constant $\alpha_s(\mu)$ with the wave function at the origin $|\psi(0)|^2_\mu$. Both are essential for hyperfine splittings and magnetic moments, yet cannot be disentangled unambiguously. To interpret discrepancies between our predicted M1 widths and experimental data, we examine the correlation between $\alpha_s$ and $|\psi(0)|^2$.

In EMS, the hyperfine interaction term $b_{q\overline{q}^\prime}$ encapsulates both $\alpha_s$ and $|\psi(0)|^2$, making their separation scale-dependent. As shown in Fig. \ref{Fig2}, increasing $\alpha_s$ reduces the effective $|\psi(0)|^2$, while heavier quark systems (e.g., charmonium) exhibit larger values consistent with more compact spatial structure. In contrast, the PM directly determines the wave function at the origin, $|\Psi(0)|^2$ (where $\Psi (0) = \frac{R_{nS}(0)}{\sqrt{4\pi}}$), by solving the Schrodinger equation. We list PM estimates of $|\Psi(0)|^2$, alongside EMS $|\psi(0)|^2$ values estimated using our inputs (Table~\ref{t1}) and $\alpha_s=0.56(6)$, in Table \ref{t10} for comparison. The $\alpha_s = 0.56(6)$ in EMS is a natural choice, as it reproduces charmonium masses in the PM to within $1\%$ of experimental values and lies well within the theoretically established ranges reported in the literature \cite{Barnes:2005pb, Deng:2016stx, Li:2010vx, Hao:2022vwt, Hao:2024ptu}. A variation of $\sim 10\%$ in $\alpha_s$ induces a roughly proportional change in $|\psi(0)|^2$. While $D$-meson estimates show rough agreement between EMS and PM, significant deviations occur for $J/\psi$. Unlike EMS, which absorbs scale-dependence into $b_{q\overline{q}^\prime}$, PM values are explicitly tied to potential-model parameters and are highly sensitive to their choice.

Despite their formal differences, both models show an inherent sensitivity to $\alpha_s$ and $|\psi(0)|^2$, which directly impact M1 transition magnetic moments and, consequently, M1 decay widths. This intrinsic sensitivity motivates our use of effective scale factors, which phenomenologically account for dynamical effects, such as higher-order QCD corrections, beyond the static quark model description.

\subsubsection{QCD corrections in PM}
\label{subPMQCD}

The scale-dependence effects discussed above suggest that the discrepancy between PM predictions and experimental results may originate from unaccounted QCD-induced effects. Accordingly, the phenomenological factor $f$ can be viewed as effectively capturing such contributions. In the literature \cite{Brambilla:2010cs, QuarkoniumWorkingGroup:2004kpm, Brambilla:2005zw}, the overestimation of widths, such as $J/\psi \to \eta_c \gamma$, is addressed through higher-order contributions from relativistic effects and the heavy-quark anomalous magnetic moment. Within an effective field theory (EFT) framework, the leading anomalous magnetic moment\footnote{The M1 matrix element is multiplied by $(1+\kappa_Q)$, where $\kappa_Q$ is the heavy-quark anomalous magnetic moment. At one loop, $\kappa_Q=\tfrac{2\alpha_{sc}(m_c)}{3\pi}+{\cal O}(\alpha_{sc}^2)$, yielding $(1+\kappa_Q)^2 \approx \big(1+\tfrac{4\alpha_{sc}(m_c)}{3\pi}\big)$ when higher powers of $\alpha_{sc}(m_c)$ are neglected \cite{Brambilla:2005zw}.}, combined with leading relativistic corrections, modifies the M1 width through the multiplicative factor \cite{Brambilla:2005zw, Brambilla:2010cs}
$$
\bigg[1+\frac{4}{3}\frac{\alpha_{sc}(m_c)}{\pi}-\frac{32}{27}\,\alpha_{sc}^2\!\big(m_c\,\alpha_{sc}\big)\bigg],
$$
where $\alpha_{sc}$ is the strong coupling evaluated at relevant scales\footnote{In EFT analyses of quarkonia, different scales govern spectroscopy and decays: spectra are controlled by potentials matched at soft/ultrasoft scales $\sim m_Q v, m_Q v^2$, whereas electromagnetic, leptonic, and hadronic decay widths involve short-distance coefficients renormalized at hard scale $\sim m_Q$. Consequently, it is common practice to use different $\alpha_s$ scales for spectra and decays \cite{Brambilla:2004jw, Brambilla:2010cs}.}.

It should be emphasized that the corrections are calculated at a chosen scale within a particular renormalization scheme \cite{Kwong:1987ak, Brambilla:2005zw}. A different choice of scale or scheme may shift the numerical size of the corrections without directly impacting the physical decay process. In practice, $\alpha_{sc}$ is evaluated at the hard scale $m_c$ for the anomalous magnetic moment, while the relativistic correction is governed by the soft momentum-transfer scale $m_c \alpha_{sc}$ \cite{Brambilla:2005zw, Brambilla:2010cs}. At leading-order, we obtain $\alpha_{sc}(m_c)=0.334$ and $\alpha_{sc}(m_c\alpha_{sc})=0.651$ using the RunDec package \cite{Chetyrkin:2000yt, Schmidt:2012az, Herren:2017osy}, in agreement with lattice determinations employed in our earlier work \cite{A:2023bxv}. Numerically, the anomalous magnetic moment enhances the width by $\tfrac{4}{3}\tfrac{\alpha_{sc}(m_c)}{\pi}\approx 0.142$, while the relativistic correction gives a sizable suppression of $-\tfrac{32}{27}\alpha_{sc}^2(m_c\alpha_{sc})\approx -0.502$. The resulting multiplicative reduction factor, $f' \simeq 0.64$, lowers $\Gamma(J/\psi \to \eta_c \gamma)$ to 1.667 keV, in good agreement with experiment. The above EFT based QCD corrections which yield $f'$, are derived for low-lying heavy quarkonium states. Nonetheless, when cautiously applied to heavy–light systems, $f'$ also improved charmed and charm-strange M1 widths, providing $\Gamma(D^{*+}\to D^+\gamma)=1.143$ keV in agreement with data as shown in column 5 of Table \ref{t9}. This indicates that $f'$ can, at least phenomenologically, provide a useful estimate of QCD corrections across the charm sector, though its validity beyond quarkonium should be interpreted with care.

From the EFT perspective, the anomalous magnetic moment originates at the hard scale ($\sim m_Q$), while relativistic corrections arise from soft ($\sim m_Q v$) and, in some cases, ultrasoft ($\sim m_Q v^2$) modes. In the weakly coupled regime ($m_Q v \gg \Lambda_{\text{QCD}}$), relevant to charmonium, these effects are perturbatively calculable. In the strongly coupled regime ($m_Q v \sim \Lambda_{\text{QCD}}$), however, they must be absorbed into Wilson coefficients and nonperturbative matrix elements within potential nonrelativistic QCD (pNRQCD) \cite{Brambilla:2004jw, Brambilla:2005zw, Brambilla:2010cs}. The sizable relativistic suppression highlights the difference between spectroscopy, where spin–dependent terms such as $V_{SS}(r)$ encode relativistic effects, and M1 decays, where additional QCD corrections to wave functions and matrix elements are essential. Finally, the close numerical agreement between the EFT reduction factor $f'$ and the phenomenological factor $f$ indicates that $f$ successfully parameterizes the dominant higher-order QCD contributions absent in static potential models, with demonstrated applicability beyond charmonium to charm-strange and heavy–light systems.

\subsubsection{Anomalous magnetic moment in EMS}
We emphasize with due caution that in PM framework, anomalous magnetic moment and relativistic corrections are rigorously established through matching to EFT formulations such as NRQCD and pNRQCD, where they arise as perturbative short-distance contributions to the chromomagnetic operator \cite{Brambilla:2005zw, Brambilla:2010cs, Pineda:2013lta}. In the present EMS approach, however, we adopt a purely phenomenological interpretation: the empirical scale factor $\Omega$ is treated as an effective analogue of the heavy-quark anomalous magnetic moment $\kappa_Q$, both encapsulating short-distance QCD dynamics that renormalize the transition magnetic moment $\mu_{V\to P}$ and hence modify M1 widths. The anomalous magnetic moment of the heavy quark, $\kappa_Q$, is described by a perturbative expansion in the strong coupling constant, $\alpha_s(m)$:
$$\kappa_Q = \kappa_Q^{(1)}\alpha_s(m) + \kappa_Q^{(2)}\alpha_s^2(m) + \kappa_Q^{(3)}\alpha_s^3(m) + \cdots$$
Here, the coefficients $\kappa_Q^{(n)}$ represent QCD loop corrections at scale $m$, with $n$ indicating the loop order. At leading-order, this reduces to  
\begin{equation}
\kappa_Q = \frac{2\alpha_s}{3\pi},
\end{equation}
which accounts for gluonic vertex corrections that enhance the heavy quark’s magnetic coupling. More precise, higher-order corrections are detailed in the work by Grozin \textit{et al.} \cite{Grozin:2007fh}. In EMS, since the hyperfine term $b_{q\bar q'}$ already encodes the dominant $\alpha_s$ and $|\psi(0)|^2$ dependence relevant to spectroscopy, the residual short-distance effects are most naturally incorporated through a phenomenological factor $(1+\kappa_Q)$. This interpretation is consistent with the argument of Fomin \textit{et al.}~\cite{Fomin:2019wuw,
Fomin:2017ltw}, that highlighted the necessity of allowing the heavy-quark $g$-factor to deviate from its canonical value for a realistic description of magnetic moments.

Accordingly, we introduce the corrected transition magnetic moment  
\begin{equation}
\mu_{V\to P}^{\rm corr} \approx (1+\kappa_Q)\,\mu_{V\to P},
\end{equation}
and calibrate $\kappa_Q$ phenomenologically in the charm sector. Utilizing $\alpha_s=0.56(6)$, we obtain $\kappa_Q \approx 0.324(57)$ at 2-loop order, corresponding to an effective magnetic coupling factor $(1+\kappa_Q) \approx 1.324(57)$, which is approximately equal to the scale factor $\Omega$. This enhancement yields a $J/\psi \to \eta_c$ transition moment of $1.053(45)~\mu_N$ and a decay width of $\Gamma(J/\psi \to \eta_c\gamma)=1.041(90)$~keV (see Fig.~\ref{Fig1}). Furthermore, this $\Omega' \sim (1+\kappa_Q)$ factor improves the prediction for $\Gamma(D^{*+}\to D^+\gamma)$ to within $\sim 7\%$ of the experimental value. Despite this improvement, the 2-loop correction still underestimates $\Gamma(J/\psi \to \eta_c\gamma)$, indicating the need of higher-order effects. Incorporating the 3-loop correction of Ref.~\cite{Grozin:2007fh} gives $(1+\kappa_Q)=1.757(197)$ and a width of $1.834(414)$~keV. While the 3-loop result agrees better with experiment and lattice QCD result for $J/\psi \to \eta_c\gamma$, the 2-loop result proves more consistent for $\Gamma(D^{*+}\to D^+\gamma)$ as shown in column 3 of Table \ref{t9}. Notably, the phenomenological factor $\Omega=1.28$ aligns well with the 2-loop estimate, suggesting that in EMS different decay channels may favor different effective corrections, with higher-order QCD contributions essential for improved accuracy. Although our use of $\kappa_Q$ is phenomenological and not derived from a systematic EFT matching, its consistent ability to improve M1 width predictions across both charmonium and heavy–light systems suggests that it provides a practical method for incorporating missing QCD corrections into the effective magnetic moment framework. In this regard, its role is analogous to the phenomenological reduction factor $f'$ previously used in the PM description (Sec.~\ref{subPMQCD}), with the key distinction that relativistic corrections are absorbed into the constituent quark masses in EMS, as discussed in Sec.\ref{EMS}.

A fundamental distinction between the PM and EMS approaches lies in the treatment of the strong coupling constant, $\alpha_s$. In PM, $\alpha_s$ is treated as an effective parameter, tuned to reproduce the charmonium spectrum. When applied to dynamical observables like anomalous magnetic moments ($\kappa_Q$), this tuned value effectively corresponds to a lower coupling ($\alpha_{sc}$) at renormalized scale. In contrast, the EMS utilizes a fixed $\alpha_s$ value, dynamically determined from the hyperfine splitting, thereby directly encoding the effective spin-spin interaction strength. Note that we adopt $\alpha_s = 0.56(6)$ for the anomalous magnetic moment calculation within the EMS, ensuring consistency with its prior use in determining $|\psi(0)|^2$. This conceptual difference has profound implications for dynamical observables. We evaluate the anomalous magnetic moment, $\kappa_{Q}(\mu)=\frac{2\alpha_{s}(\mu)}{3\pi}$, using the leading-order running coupling\footnote{Employing the known value of $\alpha_s(\mu_0)$ at a particular scale $\mu_0$, we find $\alpha_s(\mu) = \frac{\alpha_s(\mu_0)}{(1+\beta_0\alpha_s(\mu_0)~ln(\mu^2/\mu_0^2)}$ at leading-order, where $\beta_0 = \frac{1}{4\pi}(11 - \frac{2}{3}n_f)$ and $n_f$ is the number of active flavors \cite{ParticleDataGroup:2024cfk, Prosperi:2006hx}.} with initial values $\alpha_s^{\rm EMS}(1.5~\text{GeV})=0.56$ and $\alpha_{sc}^{\rm PM}(1.5~\text{GeV})=0.334$. As shown in Fig. \ref{Fig3}, while $\kappa_Q$ is stable near the charm mass, it increases significantly at lower scales, an effect more pronounced in EMS due to its larger coupling. At $\mu=1.5$ GeV, the ratio $\kappa_Q^{\rm EMS}/\kappa_Q^{\rm PM}\sim1.7$ directly reflects the ratio $\alpha_s^{\rm EMS}/\alpha_{sc}^{\rm PM}$. Thus, this factor of $\sim1.7$ is not a discrepancy but a direct consequence of the different physical roles of $\alpha_s$ in each approach. Consequently, the scale-dependence of $\alpha_s$ is a central and unavoidable factor for obtaining reliable estimates of $\kappa_Q$ and, hence, accurate M1 decay widths. 

For light mesons, where the nonrelativistic EFT hierarchy is less reliable, EMS naturally compares with VMD and chiral models. As stated before, unlike other frameworks, which use scale factors, EMS ties the short-distance coupling to the hyperfine splitting and reproduces M1 widths without ad-hoc rescaling. The EMS correction $\kappa_Q$ may be regarded as the static analogue of the one-loop anomalous magnetic moment in NRQCD, while in dynamical form factor models the same physics is embedded in the normalization and $q^2$ dependence of $g_{VP\gamma}(q^2)$. Since $g_{VP\gamma}$ and the static transition moment $\mu_{V\to P}$ are proportional up to kinematic factors, discrepancies in intrinsic magnetic moments translate directly into deviations in predicted widths, as evident from Tables~\ref{t7} and \ref{t8}. In the heavy sector, EMS employs the same $\alpha_s^{\rm EMS}$ and absorbs residual relativistic and hadronic effects into a single factor $\Omega$, offering a uniform and more constrained correction than channel-specific rescalings. This yields a transparent mapping from short-distance couplings to M1 widths, with remaining uncertainties dominated by scale choice, relativistic overlaps, and higher-order QCD effects, to be reported alongside $\Omega$.

An instructive pattern emerges when comparing uncorrected EMS and PM predictions with experimental data. The EMS framework, while reproducing light-meson M1 widths, underestimates heavy-flavor transitions because the anomalous magnetic moment is not yet included. This missing piece is naturally included in a scale factor $\Omega' \sim 1+\kappa_Q$, the analogue of the reduction factor $f'$ in PM. Both parametrize the same QCD dynamics, vertex and relativistic corrections, though absorbed at different levels: EMS introduces them explicitly as a dynamical shift, while PM (and VMD) incorporate them implicitly at the amplitude level. In this sense, $\Omega'$ in EMS and $f'$ in PM are complementary realizations of the same corrections: EMS provides a transparent, quark-level baseline tied to hyperfine splittings, whereas PM and form factor approaches achieve greater dynamical completeness. Their interplay underscores that reliable M1 predictions require both interpretability and dynamical accuracy, ultimately yielding experimentally testable benchmarks for heavy-flavor decays.

\section{Summary and Conclusions}
\label{Conclusions}
In the present work, we studied the magnetic (transition) moments and radiative M1 decay widths of ground state mesons consisting light and charm quarks. We employed the effective mass scheme to obtain the effective quark (antiquark) masses, enabling the determination of the vector meson magnetic moments. The constituent quark masses and strong hyperfine interaction terms are obtained from the experimental masses of ground state mesons in a parameter-independent approach. We have partially incorporated flavor-dependent effects through the calculation of $b_{ij}$ terms from respective flavor sectors. Therefore, the results obtained in this model-independent way ensure the robustness of our predictions. Consequently, we calculated the $V \to P$ transition moments, which are subsequently used to predict the radiative M1 decay widths of light as well as charm mesons. Furthermore, in the case of heavy flavor sector, we enhanced our predictions by incorporating a scale factor $\Omega$ (obtained by comparing the experimental coupling $g_{VP\gamma}^{Expt}$ of $D^{*+}\to D^+\gamma$ with our calculated $g_{VP\gamma}^{EMS}$) in charm meson transition magnetic moments. In addition, we predicted the $V \rightarrow P \gamma$ decay widths of charm sector mesons by employing PM to supplement our understanding of the radiative transitions. We optimized the phenomenological parameters to reliably predict bound state isoplet masses and estimate bound state spatial wave functions. Subsequently, we predicted the M1 decay widths. Furthermore, we introduced a phenomenological correction factor, $f$, estimated from the experimental decay width of $J/\psi \to \eta_{c}\gamma$, to improve our PM predictions for all decays. In addition, we estimated the total width of $D^{*0}$ and $D_s^{*+}$ mesons using our calculated M1 decay widths and experimental branching ratios. Furthermore, we incorporate leading anomalous magnetic moment and relativistic corrections to M1 decays within PM, deriving an EFT-based reduction factor $f'$ that is numerically close to the phenomenological $f$. This indicates that $f$ effectively captures higher-order QCD contributions absent in the static potential framework, with applicability across the charm sector. Similarly, in EMS, the empirical scale factor may be interpreted at the quark level as the anomalous magnetic moment correction, $\Omega' \sim (1+\kappa_Q)$, thereby encoding residual short-distance dynamics. This perspective underscores the need for a non-canonical heavy-quark $g$-factor in realistic descriptions of magnetic moments, consistent with the arguments of Fomin \textit{et al.}~\cite{Fomin:2019wuw,Fomin:2017ltw}. Based on our analysis, the following conclusions have been drawn:
\begin{itemize}
\item In the light meson sector, our EMS result $\mu_{\rho^+} = 2.440~\mu_N$ agrees well with (indirect) estimate of $\mu_{\rho^+} = 2.6(6)~\mu_N$ \cite{GarciaGudino:2015ocw} as well as other theoretical models. Furthermore, our magnetic moment results for vector mesons are consistent with various theoretical model predictions, for both light as well as heavy flavors. In the case of neutral mesons, \textit{i.e.}, $K^{*0}~\text{and}~D^{*0}$, the magnetic moment is negative in all theoretical models due to the predominant contribution from the negatively charged lighter quark (antiquark) flavor inside neutral mesons. 
\item It is worth noting that our transition moment results are larger, roughly twice those of other theoretical models. This is due to the fact that EMS uniquely derives constituent quark masses and $b_{ij}$ terms directly from meson spectra. This approach consistently yields smaller quark masses, which in turn results in larger transition magnetic moments, especially noticeable in the light quark sector. However, this eliminates the necessity for an arbitrary scale factor (particularly involving light mesons) often employed in other theoretical approaches. 
\item Our analysis conclusively demonstrates the critical role of transition mass selection in calculating transition moments, especially for light mesons. The superior agreement of our M1 decay width predictions for light mesons with experimental data validates using the arithmetic mean as the transition mass over the geometric mean. This distinction, however, diminishes in the heavy meson sector, due to the reduced contributions from strong hyperfine interactions.
\item We observe that our M1 decay width results for light mesons are consistent with the experimental values, except for $\phi \to \eta^{(\prime)}\gamma$. This exception in predictions involving $\eta^{(\prime)}$ mesons in the final states has also been observed in other theoretical works. Interestingly, our M1 decay width results are in good agreement with the predictions of theoretical models like EBM \cite{Simonis:2018rld, Simonis:2016pnh} and RPM \cite{Jena:2010zza}, although our transition moment results are larger roughly by a factor of two. We want to emphasize that none of the theoretical models, in general, has been able to consistently match experimental data on M1 decay widths for all $V \to P$ transitions in both light and heavy meson sectors. 
\item As previously discussed, our EMS calculations underestimate the experimental M1 decay widths of charm mesons by $39\%$ for $D^{*+}\to D^{+}\gamma$ and $62\%$ for $J/\psi\to \eta_{c}\gamma$. Conversely, our PM results overestimate the experimental values by $34\%$ and $66\%$ for these respective transitions. 
\item Although, there exists significant uncertainties in the experimental decay widths, \textit{i.e.}, $\sim \mathcal{O}(25\%)$, we attempted to bring improvement in our charm meson M1 decay width results through a scale factor in transition magnetic moments. Building on the model-independent relation $g_{VP\gamma} \propto \mu_{V \to P}$, we determined the scaling factor $\Omega = 1.28$ by comparing our EMS-calculated static moment with the experimentally required $g_{VP\gamma}$ for $D^{*+} \to D^+\gamma$ transition decay width. Applying this factor universally across charm sector transitions significantly improved our agreement with experimental observations. The inclusion of $\Omega$ resulted in a reduction of deviation for $\Gamma_{J/\psi \to \eta_c\gamma}$ prediction with respect to the experimental value from $62\%$ to $38\%$. Furthermore, our refined EMS results for $\Gamma_{D^{*0}\to D^{0}\gamma}$ and $\Gamma_{D^{*+}_{s}\to D^{+}_{s}\gamma}$ are consistent with theoretical predictions. 
\item Building on the PM approach's natural compatibility with heavy-quark dynamics, we further refined its predictions by introducing a flavor- and scale-dependent correction factor $f = 0.603$, determined from the $J/\psi \to \eta_c\gamma$  transition. This phenomenological adjustment, representing non-relativistic and QCD effects beyond the static potential, simultaneously improved predictions for all charm sector decays, notably achieving excellent agreement for $D^{*+} \to D^+\gamma$. Remarkably, after applying our respective scaling factors (enhancement for EMS, reduction for PM), both approaches converge to consistent results, demonstrating that systematic scale adjustments can bridge between light-quark optimized and heavy-quark tuned frameworks.  

\item The convergence of EMS and PM arises through the scale factors $\Omega'$ and $f'$, which, though introduced differently, both capture the same short-distance QCD dynamics. In PM, $f'$ incorporates anomalous magnetic moment and relativistic corrections, reducing the predicted $J/\psi \to \eta_c \gamma$ width and improving consistency across the charm sector. In EMS, the empirical factor $\Omega' \sim (1+\kappa_Q)$ provides a quark-level interpretation of missing short-distance contributions, enhancing transition moments and decay widths toward experimental values. These corrections reflect the QCD scale-dependence that links $\alpha_s$ with $|\psi(0)|^2$, explaining the discrepancies between uncorrected model predictions and data. Thus, despite originating from distinct frameworks, both EMS and PM encode complementary manifestations of the same underlying QCD effects, underscoring the role of heavy-flavor M1 transitions as precision tests of quark model dynamics.
\end{itemize} 

Despite current experimental limitations in comprehensively understanding $V \to P\gamma$ decays, our study provides valuable insights into meson magnetic properties, offering benchmarks for future investigations. While our scale-factor approach improved predictions, complete consistency across all decays remains challenging due to varying phase spaces highlighting the need for decay-specific treatments. We emphasize that higher-precision experimental data will be crucial to resolve these outstanding questions and guide theoretical refinements.

\section*{Acknowledgment}
The author RD gratefully acknowledge the financial support by the Department of Science and Technology (SERB:TAR/2022/000606), New Delhi.

\newpage

	\begin{table}[ht]
		\centering
		\captionof{table}{Magnetic moments of mesons (in units of nuclear magneton, $\mu_N$).} 
		\label{t3}
		\begin{tabular}{|c|c|c|c|c|c|c|c|c|c|c|c|}
			\hline
			\textbf{Vector} & \multicolumn{2}{c|}{\textbf{This work}} & \textbf{EBM} & \textbf{LCQSR} & \textbf{MIT} & \textbf{NJL} & \textbf{DSE} & \textbf{DSE/BSE} & \textbf{RH} & \textbf{LQCD} & \textbf{$\chi$PT} \\ \cline{2-3}
			\textbf{States} & \textbf{EMS} & \textbf{CQM} & \bf{\cite{Simonis:2018rld}} & \bf{\cite{Aliev:2019lsd, Aliev:2009gj}} & \textbf{BM \cite{Zhang:2021yul}} & \bf{\cite{Luan:2015goa}} & \bf{\cite{Bhagwat:2006pu}} & \bf{\cite{Xu:2024fun}} & \bf{\cite{Badalian:2012ft}} & \bf{\cite{Owen:2015gva, Lee:2008qf}} & \bf{\cite{Wang:2019mhm}} \\ \hline
			$\rho^{+}$  & $2.440$  & $3.068$  & $2.65$   & $2.9(5)$   & -       & $2.54$ & $2.43$  & $2.492$ & $2.37$   & $2.61(10)$ & -                        \\
			$K^{*+}$    & $2.337$  & $2.685$  & $2.35$   & $2.1(4)$   & $2.30$   & $2.26$ & $2.34$  & $2.261$ & $2.19$   & $2.81(1)$ & -                        \\
			$K^{*0}$    & $-0.299$ & $-0.383$ & $-0.23$ & $0.29(4)$ & $-0.18$ & -      & $-0.27$ & $-0.119$ & $-0.18$ & -  & -                        \\
			$D^{*0}$    & $-1.543$ & $-1.651$ & $-1.28$  & $0.30(4)$  & $-0.98$ & -      & -       & $-0.944$ & -        & -  & $-1.48 ^{+0.38} _{-0.22}$\\
			$D^{*+}$    & $1.356$  & $1.417$  & $1.13$   & $1.16(8)$  & $1.21$  & $1.16$ & -       & $1.132$ & -        & -  & $1.62 ^{+0.08}_{-0.24}$  \\
			$D^{*+}_{s}$& $1.006$  & $1.033$  & $0.93$  & $1.00(14)$ & $1.08$  & $0.98$ & -       & $1.007$ & -        & -  & $0.69^{+0.10}_{-0.22}$   \\ \hline
		\end{tabular}
	\end{table}
	\begin{table}[ht]
	\centering
	\captionof{table}{Photon momenta\footnote{Note that both $\omega$ and $k_\gamma$ are computed using Eq. \eqref{eq16} with the experimental masses of vector and pseudoscalar mesons used to calculate $\omega$, while $k_\gamma$ is calculated using theoretical masses determined from PM approach.}, $\omega$ and $k_\gamma$, for light and heavy mesons (in MeV).} 
	\label{t6}
	\begin{tabular}{|c|c|c|c|} \hline  
		\textbf{Transitions}& $\omega$ & $k_\gamma$ & \textbf{PDG \cite{ParticleDataGroup:2024cfk}} \\  \hline\hline
		$\rho^{+}\to \pi^{+}$    & $371.834$  & - & $375$ \\
		$\rho^{0}\to \pi^{0}$    & $372.654$  & - & $376$ \\
		$\rho\to \eta$           & $189.342$  & - & $194$ \\
		$\omega\to \pi$          & $379.691$  & - & $380$ \\
		$\omega\to \eta$         & $199.578$  & - & $200$  \\
		$\phi\to \eta$           & $362.519$  & - & $363$ \\
		$\phi\to \eta^{\prime}$  & $59.815$   & - & $60$ \\
		$K^{*+}\to K^{+}$        & $311.704$  & - & $309$ \\
		$K^{*0}\to K^{0}$        & $309.527$  & - & $307$ \\ \hline \hline 
		$D^{*0}\to D^{0}$        & $136.985$  & $135.840$ & $137$ \\
		$D^{*+}\to D^{+}$        & $135.683$  & $135.840$ & $136$ \\
		$D^{*+}_{s}\to D^{+}_{s}$& $138.952$  & $139.015$ & $139$ \\
		$J/\psi\to \eta_{c}$     & $110.746$  & $110.883$ & $111$ \\ \hline
	\end{tabular}
\end{table}
	\begin{table}[ht]
		\centering
		\captionof{table}{Transition moments of light mesons (in $\mu_N$).} 
		\label{t4}
		\begin{tabular}{|c|c|c|c|} \hline  
			\textbf{Transitions}& \textbf{EMS} & \textbf{EBM \cite{Simonis:2018rld, Simonis:2016pnh}} & \textbf{RPM \cite{Jena:2010zza}} \\  \hline\hline
			$\rho^{+}\to \pi^{+}$   & $1.377$  & $0.720$ & $0.695$ \\
			$\rho^{0}\to \pi^{0}$   & $1.377$  & $0.720$ & $0.694$ \\
			$\rho\to \eta$   & $2.753$  & $1.66$ & $1.948$ \\
			$\omega\to \pi$ & $4.131$  & $2.15$  & $2.069$ \\
			$\omega\to \eta$ & $0.918$  & $0.60$  & $0.588$ \\
			$\phi\to \eta$           & $0.947$ & $0.65$ & $0.608$ \\
			$\phi\to \eta^{\prime}$ & $-0.934$ & $0.82$ & $0.888$ \\
			$K^{*+}\to K^{+}$       & $1.735$  & $0.905$ & $0.882$ \\
			$K^{*0}\to K^{0}$       & $-1.936$ & $1.27$  & $1.225$ \\ \hline 
		\end{tabular}
	\end{table}
	\begin{table}[ht]
	\centering
	\captionof{table}{Transition moments of charm mesons (in $\mu_N$).} 
	\label{t5}
	\begin{tabular}{|c|c|c|c|c|c|} \hline  
		\textbf{Transitions} & \textbf{EMS} & \textbf{EBM \cite{Simonis:2018rld}} & \textbf{$\chi$PT \cite{Wang:2019mhm}} & \textbf{RPM \cite{Jena:2002is}} & \textbf{CQM \cite{Wang:2023bek}} \\  \hline\hline
		$D^{*0}\to D^{0}$         & $2.570$  & $1.80$  & $1.57$ & $1.64$ & $2.134$ \\
		$D^{*+}\to D^{+}$         & $-0.687$ & $0.418$ & $-0.34$  & $0.49$ & $-0.515$ \\
		$D^{*+}_{s}\to D^{+}_{s}$ & $-0.265$ & $0.240$ & $0.21$ & $0.25$ & - \\
		$J/\psi\to \eta_{c}$      & $0.795$  & $0.634$ & -       & - & - \\	\hline 
	\end{tabular}
\end{table}
	\begin{table}[ht]
		\centering
		\captionof{table}{Radiative M1 decay widths of light mesons (in keV).} 
		\label{t7}
		\begin{tabular}{|c|c|c|c|c|c|c|c|c|c|} \hline  
			\multirow{2}{*}{\textbf{Transitions}}& \multirow{2}{*}{\textbf{EMS}} & \textbf{EBM} & \textbf{NJL} & \textbf{DSE} & \textbf{RPM} & \textbf{LFHM} & \textbf{LFQM} & \textbf{PM(AP1)} & \textbf{PDG} \\ 
			&  & \bf{\cite{Simonis:2018rld, Simonis:2016pnh}} & \bf{\cite{Deng:2013uca}} & \bf{\cite{Maris:2002mz}} & \bf{\cite{Jena:2010zza}} & \bf{\cite{Ahmady:2020mht}} & \bf{\cite{Choi:1997iq}} & \bf{\cite{Bonnaz:2001aj}} & \bf{\cite{ParticleDataGroup:2024cfk}} \\ \hline\hline
			$\rho^{+}\to \pi^{+}\gamma$   & $67.345$ & $76$   & $21.9$ & $52$ & $68.12$ & $66.37(7)$ & $76$               & $60.41$ & $67.59(7.59)$    \\
			$\rho^{0}\to \pi^{0}\gamma$   & $67.792$ & $76$   & $43.9$ & $52$ & $68.85$ & $66.37(7)$ & -                  & $60.64$ & $69.28(11.80)$   \\
			$\rho\to \eta\gamma$      & $35.544$  & $55.7$ & -      & - & $71.10$ & $45.73(6.16)$ & $59$               & $60.63$ & $44.22(3.10)$   \\
			$\omega\to \pi\gamma$ & $645.347$& $694$  & $866$  & $479$ & $646.11$& $625.38(66.03)$ & $730(1.3)$         & $571.79$& $724.78(25.83)$  \\
			$\omega\to \eta\gamma$    & $4.625$  & $7.96$ & -      & - & $7.57$  & $4.93(75)$ & $6.9(3)$           & $7.72$  & $3.91(35)$ \\
			$\phi\to \eta\gamma$          & $29.524$  & $54.7$ & -      & - & $48.80$ & $67.63(9.21)$ & $49.2(1.6)$        & $44.12$ & $55.28(1.03)$  \\
			$\phi\to \eta^{\prime}\gamma$ & $0.129$  & $0.384$& -      & - & $0.47$  & $0.36(5)$ & $0.36(1)$          & $0.32$  & $0.26(1)$ \\
			$K^{*+}\to K^{+}\gamma$       & $62.988$ & $68$   & $13.5$ & - & $63.62$ & $81.20(20.66)$ & $79.5$             & $104.46$& $50.37(4.69)$    \\
			$K^{*0}\to K^{0}\gamma$       & $76.803$ & $134$  & $31.3$ & - & $123.40$& $122.02(10.49)$ & $124.5$            & $116.41$& $116.36(10.01)$  \\ \hline 
		\end{tabular}
	\end{table}
	\begin{sidewaystable}[ht]
		\centering
		\captionof{table}{Radiative M1 decay widths of charm mesons (in keV).} 
		\label{t8}
		\begin{tabular}{|c|c|c|c|c|c|c|c|c|c|c|c|c|c|c|c|} \hline  
			\multirow{2}{*}{\textbf{Transitions}}& \multicolumn{2}{c|}{\textbf{This work}} & \textbf{EBM} & \textbf{LCQSR} & \textbf{NJL} & \textbf{LQCD} & \textbf{$\chi$PT} & \textbf{LFQM} & \textbf{QCDSR} & \textbf{PM(AP1)} & \textbf{RIQ} & \textbf{RQM} & \textbf{CM} & \textbf{GI} & \textbf{PDG} \\ \cline{2-3}
			& \textbf{EMS} & \textbf{PM} & \bf{\cite{Simonis:2018rld}} & \bf{\cite{Pullin:2021ebn}} & \bf{\cite{Deng:2013uca}} & \bf{\cite{Meng:2024gpd, Colquhoun:2023zbc, Becirevic:2009xp}} & \bf{\cite{Wang:2019mhm}} & \bf{\cite{Choi:2007se}} & \bf{\cite{Lu:2024tgy}} & \bf{\cite{Bonnaz:2001aj}} & \bf{\cite{Priyadarsini:2016tiu}} & \bf{\cite{Ebert:2002xz, Ebert:2002pp}} & \bf{\cite{Cheung:2015rya, Cheung:2014cka}} & \bf{\cite{Godfrey:2015dva}} & \bf{\cite{ParticleDataGroup:2024cfk}} \\ \hline\hline
			$D^{*0}\to D^{0}\gamma$         & $11.731$ & $35.506$ & $22.9$ & $27.83^{+9.23}_{-9.50}$ & $19.4$ & $27(14)$ & $16.2^{+6.5}_{-6.0}$  & $21.0(3)$ & $1.74^{+0.40}_{-0.37}$ & $41.74$ & $26.509$ & $11.5$ &  $22.7^{+2.1}_{-2.2}$ & $106$ & -            \\
			$D^{*+}\to D^{+}\gamma$         & $0.815$  & $1.785$  & $1.19$ & $0.96^{+0.58}_{-0.62}$ & $0.7$ & $0.8(7)$ & $0.73^{+0.7}_{-0.3}$  & $0.96(2)$ & $0.17^{+0.08}_{-0.07}$ & $3.58$ & $0.932$ & $1.04$ & $0.9^{+0.3}_{-0.2}$ & $10.8$ & $1.33(33)$ \\
			$D^{*+}_{s}\to D^{+}_{s}\gamma$ & $0.130$  & $0.284$ & $0.43$ & $2.36^{+1.49}_{-1.41}$ & $0.09$ & $0.055(5)$ & $0.32^{+0.3}_{-0.3}$  & $0.17(1)$ & $0.029^{+0.009}_{-0.008}$ & $0.31$ & $0.213$ & $0.19$ & $3.53$ & $1.03$ & -          \\
			$J/\psi\to \eta_{c}\gamma$      & $0.594$  & $2.605$ & $1.56$ & - & - & $2.219(35)$ & -                     & $1.65(5)$ & - & $1.87$ & $2.321$ & $1.05$ & - & - & $1.57(37)$ \\	\hline 
		\end{tabular}
	\end{sidewaystable}

\begin{table}[ht]
\centering
\captionof{table}{Improved radiative M1 decay widths (in keV) using phenomenological factors in EMS ($\Omega$, $\Omega'$) and PM ($f$, $f'$) approaches.} 
\label{t9}
    \begin{tabular}{|c|c|c|c|c|c|c|c|} \hline  
	\textbf{Transitions} & \textbf{EMS} \bm{$(\Omega)$} & \textbf{EMS} \bm{$(\Omega')$}\footnote{$\Omega'$ represents the anomalous magnetic moment correction calculated at 2-loop order.} & \textbf{PM} \bm{$(f)$} & \textbf{PM} \bm{$(f')$}\footnote{$f'$ encodes anomalous magnetic moment and relativistic correction computed with $\alpha_{sc}$.} & \textbf{PDG \cite{ParticleDataGroup:2024cfk}}   \\ \hline\hline
	$D^{*0}\to D^{0}\gamma$          & $19.122$ & $20.564(1.764)$ & $21.41$ & $22.724$ & - \\
	$D^{*+}\to D^{+}\gamma$          & $1.328$ & $1.429(123)$ & $1.076$ & $1.143$ & $1.33(33)$ \\
	$D^{*+}_{s}\to D^{+}_{s}\gamma$  & $0.212$ & $0.228(20)$ & $0.171$ & $0.182$ & - \\
	$J/\psi\to \eta_{c}\gamma$       & $0.973$ & $1.041(90)$ & $1.571$ & $1.667$ & $1.57(37)$\footnote{Used as input to calculate $f$.}  \\ 
     & & [$1.834(414)$]\footnote{Numerical result with 3-loop correction in $\kappa_Q$.} &  &  &   \\ \hline 
    \end{tabular}
\end{table}

\begin{table}[]
    \centering
    \caption{Square of the wave function at the origin (in GeV${}^3$) of charm sector mesons in EMS and PM.}
    \label{t10}
    \begin{tabular}{|c|c|c|}
    \hline
         \textbf{States} &  $\bm{|\psi(0)|^2}$ \textbf{(EMS)}\footnote{Extracted using corresponding inputs (Table \ref{t1}) and with $\alpha_s = 0.56(6)$.} & $\bm{|\Psi(0)|^2}$ \textbf{(PM)}\\
         \hline\hline
        $c\overline{u}, c\overline{d} : D^*, D$ &  $0.011(1)$ & $0.015$ \\
        $c\overline{s} : D^{*+}_{s}, D^{+}_{s}$ &  $0.018(2)$ & $0.031$ \\
        $c\overline{c} : J/\psi, \eta_{c}$ &  $0.046(5)$ & $0.118$ \\
        \hline
    \end{tabular}
\end{table}

\end{document}